\documentclass[12pt]{article}

\usepackage{amsfonts}
\usepackage{amsmath}
\usepackage{amssymb,mathrsfs}
\usepackage{amsthm}

\usepackage{amsmath}
\usepackage{amsfonts}

\newlength{\cellwidth}
\usepackage{tabularx}
\usepackage{multirow}
\usepackage{array}
\newcommand{\specialcell}[2][c]{%
  \begin{tabular}[#1]{@{}p{\cellwidth}@{}}#2\end{tabular}}

\newcolumntype{P}[1]{>{\centering\arraybackslash}p{#1}}
\newcolumntype{M}[1]{>{\centering\arraybackslash}m{#1}}

\def\vs{\vspace{5mm}}

\newtheorem{theorem}{Theorem}

\newtheorem{Proposition}{Proposition}
\newtheorem{lemma}{Lemma}

\newcommand{\bm}[1]{\mbox{\boldmath $#1$}}

\newcounter{mnotecount}

\newcommand{\mnotex}[1]
{\protect{\stepcounter{mnotecount}}$^{\mbox{\footnotesize $\bullet$\themnotecount}}$ 
\marginpar{
\raggedright\tiny\em
$\!\!\!\!\!\!\,\bullet$\themnotecount: #1} }

\def\defi{:=}

\def\s{s}

\def\H{{\cal H}}
\def\M{{\cal M}}
\def\C{{\cal C}}
\def\I{{\cal I}}
\def\F{{\cal F}}
\def\X{{\cal X}}
\def\Y{{\cal Y}}

\def\W{{\cal W}}
\def\U{{\cal U}}
\def\V{{\cal V}}

\def\mbar{\overline{m}}

\def\grad{\mbox{grad}}

\def\be{\begin{equation}}
\def\ee{\end{equation}}
\def\bea{\begin{eqnarray}}
\def\eea{\end{eqnarray}}
\def\bean{\begin{eqnarray*}}
\def\eean{\end{eqnarray*}}

\def\scri{\mathscr{J}}

\def\ii{\mbox{\rm I\!I}}

\begin{document}

\title{Spacetime characterizations of 
$\Lambda-$vacuum
metrics with a null Killing 2-form 
}
\author{Marc Mars$^{1}$ and Jos\'e M. M. Senovilla$^{2}$ \\
\\
$^{1}$ Instituto de F\'{\i}sica Fundamental y Matem\'aticas, Universidad de Salamanca\\
Plaza de la Merced s/n, 37008 Salamanca, Spain\\
marc@usal.es\\
\\
$^{2}$ F\'{\i}sica Te\'orica, Universidad del Pa\'{\i}s Vasco, \\
Apartado 644, 48080 Bilbao, Spain \\ 
josemm.senovilla@ehu.es}
\date{}
\maketitle

\begin{abstract} 
An exhaustive list of four-dimensional $\Lambda$-vacuum spacetimes admitting a Killing vector $\xi$ whose self-dual Killing two-form $\F_{\mu\nu}$ is null is obtained assuming that the self-dual Weyl tensor is proportional to the tensor product of $\F_{\mu\nu}$ by itself.
Our analysis complements previous results \cite{Mars1,MS} concerning the case with non-null $\F_{\mu\nu}$. 
We analyze both cases with $\Lambda$ vanishing or not. In the latter case we prove that $\Lambda<0$ must hold necessarily, and we find a characterization of the Einstein spacetimes conformal to pp-waves. In the former case we obtain spacetime characterizations of vacuum plane waves and of the stationary vacuum Brinkmann spacetimes. 
At the light of the full set of results, old and new, we reformulate the case with non-null $\F_{\mu\nu}$ and $\Lambda =0$. We finally present a table collecting the results for both null, and non-null, $\F_{\mu\nu}$.
\end{abstract}

\section{Introduction}

Finding properties that  locally characterize spacetimes 
is a relevant problem 
both in geometry and in physics as it provides us, among other things, with tools
to approach uniqueness results under suitable global assumptions and
with methods to describe quantitative 
deviations from a given spacetime.
Characterizations involving the vanishing of a tensorial object
are particularly relevant in this respect. The classic and certainly most 
important example is the vanishing of the Riemann tensor that characterizes locally flat metrics. 
In General Relativity, the Kerr spacetime and its generalization
with non-zero cosmological constant $\Lambda$ and/or NUT parameter  (the so-called
Kerr-NUT or Kerr-NUT-(A) de Sitter spacetimes) play a relevant role
and substantial effort has been devoted to characterizing them locally.
A number of results along these lines are known 
involving various properties, such as for instance separability of the
Hamilton-Jacobi equation or, more geometrically, the existence of a closed
conformal Killing-Yano tensor in $\Lambda$-vacuum spacetimes
(in arbitrary dimensions) 
\cite{Krtous, Houri_a, Houri_b}. Characterizations within the class 
of $\Lambda$-vacuum spacetimes of Petrov type D with
shear-free and geodesic principal null congruences have been
obtained by Ferrando and S\'aez \cite{FerrandoSaez}.

In the case when the spacetime admits a local isometry with Killing 
generator  $\xi$, the Kerr metric was found \cite{Mars1} to be uniquely
characterized among stationary and asymptotically flat vacuum
spacetimes by the property that the self-dual Weyl tensor
and the self-dual Killing two-form satisfy a suitable alignment
condition, (\ref{C=FF1}) below. This condition was 
motivated by previous results of Simon
\cite{Simon} who characterized the Kerr metric in terms
of objects defined in the Killing vector quotient space.
The condition of asymptotic  flatness was used in \cite{Mars1}
only to fix the values of two constants and to ensure
that the Killing vector was timelike on at least one point. This leads
to a local characterization of the Kerr spacetime in terms of the 
vanishing of a suitable complex tensor (see Theorem 1 in \cite{Mars2}, which 
however needs amendment as the condition that $\xi$ is timelike
at some point was overlooked; see below for a detailed discussion of this
issue).
Such characterization has been
used successfully \cite{Ionescu1, Ionescu2}
to prove a uniqueness theorem of rotating
black holes under suitable smallness  assumptions (or appropriate
boundary conditions at the bifurcation surface) without assuming that 
the spacetime is analytic.
It has also been used to 
characterize initial data for the Kerr metric in
\cite{ParradoJuan}  and to define a quality factor
measuring the deviation of a given metric admitting a Killing vector
with respect to the Kerr metric \cite{ParradoJose}.

This local characterization of Kerr (or of Kerr-NUT, when the conditions
on the constants are relaxed \cite{Mars3}) has been  extended
\cite{MS} to the case of $\Lambda$-vacuum spacetimes. The full class of 
spacetimes satisfying the Einstein field equations with a cosmological
constant $\Lambda$ (of any value, including zero) and admitting a Killing vector
such that the alignment condition (\ref{C=FF1}) holds was studied in detail
by exploiting an interesting underlying conformal submersion that arises
naturally. In particular the Kerr-NUT-(A) de Sitter spacetime
was identified invariantly within this class. The results in \cite{MS} have been
used recently \cite{MPSS, MPS} to characterize the Kerr-NUT-de Sitter spacetimes (and other $(\Lambda>0)$-vacuum spacetimes) in terms of their Killing data
at past or future null infinity. 

All these characterizations of spacetimes in terms of (\ref{C=FF1}) made the
assumption that the square $\F^2:=\F_{\alpha\beta}\F^{\alpha\beta}$ of the self-dual Killing form $\F_{\alpha\beta}$
(see below for definitions) is non-zero somewhere ---inside the set
$\{ Q \neq 0 \}$. Then $\F^2$ is automatically non-zero everywhere and the
spacetime is of Petrov type D at every point. Given the relevance of
the spacetimes that arise under this assumption, it is most natural to
ask what happens in the complementary case when
$\F^2$ is identically zero. This is the problem we analyze in this paper.
By doing so we complete the local characterization of all
$\Lambda$-vacuum spacetimes admitting a Killing vector and satisfying
the alignment condition (\ref{C=FF1}). The spacetimes thus
characterized turn out to be, in the case $\F^2=0$, physically relevant too.  This reinforces
 the idea that the geometric condition (\ref{C=FF1}) is both
physically and geometrically meaningful.

The vanishing of $\F^2$ means that the self-dual
Killing form is null, and this introduces a privileged  null
direction $k$  defined as the wave vector of $\F_{\alpha\beta}$.
Our analysis exploits the simultaneous existence of $\xi$ and $k$ in 
the spacetime and extracts
relevant geometric information  that
plays a crucial role in determining the local form of the metric. 
Our main result is summarized
in Theorem \ref{main} of Section \ref{mainSect}. The classification splits into
two cases depending on whether the cosmological constant vanishes or not.
In the first case the spacetime is either locally isometric   to
 the vacuum plane waves (when $\xi$ and $k$ are orthogonal) or to the
vacuum stationary Brinkmann spacetimes (when they are not). Since the first class
always admits at least a five-dimensional Killing algebra and the second
one also admits more than one linearly independent
Killing vector, it may (and it does) occur that the intersection of both cases
is non-empty. We identify this intersection as the class of vacuum irreducible locally
symmetric spacetimes, which depend on a single parameter.
The case of non-zero cosmological constant requires $\Lambda <0$
necessarily and leads uniquely to the so called Siklos spacetime.

Since this paper completes the classification of $\Lambda$-vacuum spacetimes
admitting a Killing vector field $\xi$ such that (\ref{C=FF1}) holds, we include in
this paper a table that summarizes all these results. Before doing
so
 we show that the case $\{\F^2\neq 0 \}$, which in \cite{MS} was
split in three disjoint classes, can be rearranged for the case with $\Lambda=0$ into two disjoint
classes: (i) the so-called gen-Kerr-NUT, which corresponds to the
Kerr spacetime with NUT charge together with its generalizations from 
spherical to plane or hyperbolic surfaces, and (ii)  the
so-called type D vacuum Kundt. The former class is written 
in a coordinate system that covers all cases at once, including the
situation where the Killing $\xi$ is hypersurface orthogonal. This
rearrangement in the case $\{ \Lambda=0, \F^2 \neq 0\}$
is given in Theorem \ref{vacuumF2} and allows us to discuss
in detail the omission in the statement of Theorem 1 in \cite{Mars2} 
already mentioned above. Indeed, the type D vacuum Kundt class can be singled out by
the  property that the Killing vector $\xi$ is orthogonal everywhere
to the
two null eigendirections of $\F_{\alpha\beta}$. Thus, when
the Killing $\xi$ is timelike at one point the type D vacuum Kundt class 
gets automatically excluded. Given that 
\cite{Mars1} dealt with spacetimes where $\xi$ is timelike somewhere, this
case did not show up in the analysis. The problem was that 
the quotation made in Theorem 1 of \cite{Mars2} of the 
results in \cite{Mars1} did not include this assumption.
We amend the statement 
in Theorem \ref{corrected} by including all necessary assumptions
and also give an independent proof based
on the splitting in two classes presented in Theorem \ref{vacuumF2}.

The paper is organized as follows. In Section 
\ref{sec:preliminaries} we present our basic assumptions and find a number
of identities that will play a role later. In Section \ref{Lambda0} we 
study the case with vanishing cosmological constant. After proving 
in Proposition \ref{PropLambda0} that, with a suitable choice of
scale, $k$ is a parallel vector field with constant scalar 
product with $\xi$,
the spacetime is immediately shown to be a Brinkmann space
and the theory of Kerr-Schild vector fields \cite{CHS} allows us to determine
the most general form of the Killing vector $\xi$,
as well as the field equations that the remaining metric function 
must satisfy.
At this point, it becomes necessary to split the 
analysis  depending on whether the (constant) scalar product between
$\xi$ and $k$ vanishes or not, thus leading to the spacetimes identified
in Theorem \ref{main} when $\Lambda=0$. In Section \ref{Lambdaneq0} we show
in Proposition \ref{Prop2} that $\Lambda <0$ necessarily and that the
horizons of $\xi$ (which is null, hypersurface orthogonal and nowhere zero 
in this case) have cross sections of constant negative curvature. From 
here a local coordinate system is constructed in terms of
three geometrically defined scalars and a choice of  hypersurface
transversal to the horizons. The metric in this coordinate system
is simple enough so that the alignment condition (\ref{C=FF1}) can
be imposed and solved immediately. The paper closes with Section
\ref{mainSect} where our main result and the two theorems
mentioned above are stated and proven. Finally we summarize
the full local classification of spacetimes satisfying (\ref{C=FF1}),
for any value of $\Lambda$ and any value of $\F^2$, in Table \ref{Summary}.

\section{Preliminaries}
\label{sec:preliminaries}

Throughout this paper, a spacetime $(\M,g)$ is 
a smooth, orientable four-dimensional, connected
manifold endowed with a metric $g$ of
Lorentzian signature (at least $C^3$). We assume further that the spacetime
is oriented and time oriented. The Levi-Civita covariant
derivative of $g$ is denoted by $\nabla$ and the volume form
by $\eta_{\alpha\beta\gamma\delta}$. From now on a tensor field will be called smooth iff
it is $C^3$.

Our basic assumptions are
\begin{enumerate}
\item $(\M,g)$ admits a Killing vector field $\xi$ with non-identically
vanishing
self-dual {\rm Killing form} $\F_{\alpha\beta}$, defined by
\begin{align*}
\F_{\alpha \beta} \defi F_{\alpha\beta} + i F^{\star}_{\alpha\beta} , 
\hspace{1cm} 
F_{\alpha\beta} = \nabla_{\alpha} \xi_{\beta}
\end{align*}
where $\star$ is the Hodge dual operator. $F_{\alpha\beta}$ is a two-form
by the Killing equations and $\F$ is self-dual, i.e. it satisfies
$\F_{\alpha \beta}^\star =-i \F_{\alpha \beta}$.
\item The two-form $F_{\alpha \beta} $ is singular, or {\em null}, that is to say, it
satisfies
\begin{align}
\F^2 \defi \F_{\alpha\beta} \F^{\alpha\beta} =0.\label{fsquare=0}
\end{align}
\end{enumerate}
Any self-dual two-form satisfies the algebraic identities
(see e.g. \cite{Israel1970})
\begin{align*}
\F_{\mu\rho}\F_{\nu}{}^\rho = \frac{1}{4} \F^2 g_{\mu\nu}, 
\quad \quad \quad \quad
\F_{\mu\rho}\overline\F^{\mu\rho} =0, 
\end{align*}
where overbar denotes complex conjugation. Hence,
Condition (ii) can be equivalently stated as
\begin{align}
\F_{\mu\rho}\F_{\nu}{}^\rho = 0, \label{nilpotent}
\end{align}
so that the self-dual Killing two-form is nilpotent. This entails, in particular, that $F$ and $F^{\star}$ are simple 2-forms.

\subsection{Null eigenvector}
Assuming that (\ref{nilpotent}) holds at a given point $p\in \M$ where $\F$ is non-zero  it follows the existence of a unique real null 1-form $\bm{k}|_{p}\in \Lambda^{1}_{p}$ such that
$$
\bm{k} \wedge \F|_p=0 
$$
which is also equivalent to $k^{\mu}\F_{\mu\rho}|_p=0$. This implies that $\F|_{p}$ takes the simple form $\F|_{p}=\bm{k}\wedge \bm{V}|_{p}$ for some non-zero complex 
vector
$V |_p = X|_p + i Y |_p$ satisfying $g (k |_p, V|_p ) =0$ and
$g(V|_p, V|_p)=0$ (i.e.
$g (X|_p, Y|_p)  =0, g(X|_p, X|_p) = g (Y|_p, Y|_p) > 0$).
These vectors are defined uniquely up
to scaling and shift transformations, given by $\widetilde{k} |_p = a k|_p$, 
$\widetilde{V}|_p = 
a^{-1} ( V|_p + b k |_p)$, with $a\in \mathbb{R} \setminus \{ 0 \}$,
$b\in \mathbb{C}$.  The null direction $\langle k|_p \rangle$ is algebraically
characterized as the unique real direction lying in the kernel of
$\F_{\alpha\beta}$, and it is usually called the wave vector, or the {\it null eigenvector}, of $\F$.

Assume now that $\F$ is null and nowhere zero
on an open neighbourhood $\U$, then the decomposition
\begin{eqnarray}
\F_{\alpha\beta} = k_{\alpha} V_{\beta} - k_{\beta} V_{\alpha} \label{Fnull_a}
\end{eqnarray}
holds on $\U$ for smooth vector fields  $k$ and $V$ \cite{Israel1970}. We show
this explicitly for completeness. 
From
time orientability there exists a unit, future directed timelike
vector field $u$ on $\U$. The scaling and shift freedom at $p \in \U$ can
be uniquely fixed by requiring 
$g(u |_p, \widetilde{k}|_p ) = -1$
and $g(u|_p, \widetilde{V}|_p) = 0$. $\widetilde{V}$ 
is smooth on $\U$ because $\F_{\alpha\beta} u^{\beta} = \widetilde{V}_{\beta}$ 
and the left hand-side is smooth. $\widetilde{k}$ is also smooth 
on $U$ because it is  uniquely characterized by 
$g(\widetilde{k},\widetilde{k})=0, g(\widetilde{k},\widetilde{V})=0, 
g(\widetilde{k},u)=-1$. Now, the
the most general pair of vector fields $\{ k,V\}$ for which 
(\ref{Fnull_a}) holds is given by
$k' = a^{-1} \widetilde{k}$
and $V = a  \widetilde{V} - (b/a) \widetilde{k}$ where $a,b$ are smooth functions
on $\U$ with $b$ complex and $a$ nowhere vanishing. By taking
$a^{-1}$ as the positive square root of
$g (\widetilde{V},\widetilde{\overline{V}})$ (which is positive everywhere)
and defining 
$m \defi a \widetilde{V}$ 
we conclude that any 
nowhere zero
self-dual, null two-form $\F$ on an open set $\U \subset \M$
admits the following decomposition in terms of smooth $k'$ and $m$: 
\begin{eqnarray}
\F_{\alpha\beta} = k'_{\alpha} m_{\beta} - 
m_{\alpha} k'_{\beta}, \quad
\quad g (k',k')= 
g (k',m)= g (m,m)=0, 
\quad \quad g (m, \mbar )=1. \label{nullF}
\end{eqnarray}
The complex null vector field $m$ is defined up to
$m\rightarrow m + B k^{\prime}$, with $B$ an arbitrary
smooth complex function on $\U$.

\subsection{Ernst one-form and ``energy-momentum'' of $\F$.}
The Ernst one-form is the complex one-form defined by
\be 
\chi_{\beta} \defi 2 \xi^{\alpha} \F_{\alpha\beta} \label{defchi} 
\ee
and it is automatically orthogonal to $\xi$, i.e. 
$\xi^\beta\chi_\beta =0$. In the null case it also 
satisfies the identity
\be
\chi^\rho \F_{\rho\mu} =0 \label{propchi}
\ee
as a direct consequence of (\ref{nilpotent}). In fact $\chi$
can be explicitly written in terms of $k$ and $m$ as
\begin{align}
\chi_\mu = 2(\xi^\rho k'_\rho)m_\mu - 2(\xi^\rho m_\rho) k'_\mu . \label{chinull}
\end{align}
Let $N \defi -g(\xi,\xi)$
be minus the square norm of the Killing. The real part of 
the Ernst form is directly linked to $N$ by
$$
\chi_\mu +\overline\chi_\mu = 4\xi^\rho F_{\rho\mu}=-4\xi^\rho \nabla_\mu 
\xi_\rho =2\nabla_\mu N
$$
so that we can always write
\be
\chi_\mu =\nabla_\mu N +i \omega_\mu \label{DN+twist}
\ee
for some real one-form $\omega$ called the twist of $\xi$ and
which vanishes if and only if $\xi_{\alpha}$ is a 
hypersurface orthogonal one-form. 
We will also use the symmetric tensor
\begin{align}
t_{\mu\nu}\defi \frac{1}{2} \F_{\mu\rho}\overline\F_{\nu}{}^\rho =\frac{1}{2} k'_\mu k'_\nu \label{emt}
\end{align}
which is nothing but  the ``energy-momentum tensor'' of the 2-form $F_{\mu\nu}$.
$\xi$ being a Killing vector, it follows that $\pounds_{\xi}
\F_{\alpha\beta}=0$ and thus
$\pounds_{\xi} t_{\alpha\beta}=0$, 
so that (\ref{emt}) implies
\be
\pounds_\xi k'_\mu =0 \label{liek'}.
\ee



\subsection{$\Lambda$-vacuum}
Throughout this paper we will assume that the space-time satisfies Einstein's field equations for  vacuum with a (possibly vanishing)
cosmological constant $\Lambda$, that is to say, 
\be
R_{\alpha\beta} = \Lambda g_{\alpha\beta} \label{ric}
\ee
where $R_{\alpha\beta}$ is the Ricci tensor (our sign conventions follow
e.g. \cite{Wald}). Such spacetimes will be called $\Lambda$-vacuum. 
In the specific case that $\Lambda=0$ we will simply say vacuum or, also,
Ricci-flat. The Riemann tensor of $\Lambda$-vacuum spacetimes
decomposes as
\be
R_{\alpha\beta\lambda\mu}=C_{\alpha\beta\lambda\mu}+\frac{\Lambda}{3} \left(g_{\alpha\lambda}g_{\beta\mu}- g_{\alpha\mu}g_{\beta\lambda}\right) \label{RiemLam}
\ee
where $C_{\alpha\beta\mu\nu}$ is the Weyl tensor. We shall denote the (right)
self-dual of $C_{\alpha\beta\mu\nu}$ by
\begin{align*}
\C_{\alpha\beta\mu\nu} \defi C_{\alpha\beta\mu\nu} + i
C^{\star}_{\alpha\beta\mu\nu}, \quad \quad
C^{\star}_{\alpha\beta\mu\nu} \defi \frac{1}{2} \eta_{\mu\nu\rho\sigma}
C_{\alpha\beta}^{\phantom{\alpha\beta}\rho\sigma}.
\end{align*}
The algebraic properties of the Weyl tensor (i.e. vanishing trace and
first Bianchi identity) imply that $\C_{\alpha\beta\mu\nu}$ 
can be defined
equivalently using the left self-dual of $C_{\alpha\beta\mu\nu}$ (with
obvious notations) and that the following properties hold
\begin{align}
\C^{\alpha}_{\phantom{\alpha}\beta\alpha\mu} =0, \quad \quad
\C_{\alpha[\beta\mu\nu]} =0,
\label{syms}
\end{align}
where square brackets denote antisymmetrization.
The second Bianchi identities can be rewritten as
\be
\nabla^\alpha \C_{\alpha\beta\mu\nu}=0
\label{divfree}
\ee
and the standard identity for Killing vectors
$\nabla_\beta\nabla_\lambda\xi_\mu =\xi^\alpha R_{\alpha\beta\lambda\mu}$ becomes,
in the language of complex self-dual objects,
\be
\nabla_{\mu} \F_{\alpha\beta} = \xi^{\nu} \left ( \C_{\nu \mu \alpha\beta} +
\frac{4 \Lambda}{3}  \I_{\nu \mu\alpha\beta} \right ), \label{nablaF}
\ee
where 
$$
{\cal I}_{\nu\mu\alpha\beta} = \frac{1}{4} \left ( g_{\nu\alpha}
g_{\mu\beta} - g_{\nu \beta} g_{\mu\alpha} + i \eta_{\nu\mu\alpha\beta} \right ) 
$$
is the canonical metric in the space of complex self-dual 2-forms,
i.e. a symmetric by pairs ${\cal I}_{\nu\mu\alpha\beta}={\cal I}_{\alpha\beta\nu\mu}$ and satisfying  ${\cal I}_{\nu\mu\alpha\beta} \W^{\alpha\beta} = \W_{\nu\mu}$ for any self-dual 2-form $\W$.
Note that the trace of ${\cal I}$ is
\begin{align}
{\cal I}^{\alpha}_{\phantom{\alpha}\mu\alpha\beta} = \frac{3}{4} g_{\mu\beta}.
\label{trace}
\end{align}
Taking the covariant derivative in (\ref{defchi}) one finds
\begin{align}
\nabla_{\mu} \chi_{\beta} &= 2 (\nabla_{\mu} \xi^{\alpha}) \F_{\alpha\beta}
+ 2 \xi^{\alpha} \nabla_{\mu} \F_{\alpha\beta} \nonumber \\
& = ( \F_{\mu}^{\phantom{\mu}\alpha} +
\overline{\F}_{\mu}^{\phantom{\mu}\alpha}) \F_{\alpha\beta} 
+ 2 \xi^{\alpha} \xi^{\nu} \left ( \C_{\nu\mu\alpha\beta} 
+ \frac{4 \Lambda}{3} {\cal I}_{\nu\mu\alpha\beta} \right ) 
\nonumber \\
& = -k'_\mu k'_\beta+2\xi^\nu\xi^\alpha \C_{\nu\mu\alpha\beta}
-\frac{2\Lambda}{3}\left(Ng_{\mu\beta} +\xi_\mu \xi_\beta \right), \label{derchi}
\end{align}
where in the second equality we used (\ref{nablaF}) and in the
third equality we used 
the nilpotency (\ref{nilpotent}) and expression
(\ref{emt}). Hence, the Ernst one-form in $\Lambda$-vacuum spacetimes is closed
(see e.g. \cite{Mars3})
\be
\nabla_{[\mu}\chi_{\beta]} =0. \label{chiclosed}
\ee
Immediate consequences of (\ref{nablaF}), combined with (\ref{syms}) and
(\ref{trace}),  are
\begin{align}
\nabla_{[\mu} \F_{\alpha\beta]}=i  \frac{\Lambda}{3} \xi^{\nu}
\eta_{\nu\mu\alpha\beta}, \quad \quad
\nabla^\alpha\F_{\alpha\beta} =-\Lambda \xi_\beta \,,\label{maxwell}
\end{align}
while taking the derivative of the null condition
(\ref{fsquare=0}) and using
(\ref{nablaF}) gives
\begin{align}
2 \xi^{\nu} \F^{\alpha\beta} \C_{\nu\mu\alpha\beta}
+ \frac{4 \Lambda}{3} \chi_{\mu}=0\,. \label{nablaFsq}
\end{align}

As a preliminary result to be used later, we have
\begin{lemma}\label{lem:F=0}
If the self-dual Killing 2-form $\F$ vanishes on a non-empty
open neighbourhood $\U\subset \M$ of a $\Lambda$-vacuum spacetime, then $\Lambda =0$ necessarily.
\end{lemma}
\proof This follows immediately from (\ref{maxwell}).

\subsection{Alignment of $\C$ and $\F$}
The self-dual $\C_{\alpha\beta\lambda\mu}$ defines an eigenvalue problem acting on self-dual 2-forms. The content of this problem leads to the important Petrov classification of the Weyl tensor \cite{Exact}. In the case under consideration, the space-time has a distinguished self-dual 2-form, the Killing 2-form $\F_{\mu\nu}$. It seems natural to ask what are the implications of $\F_{\mu\nu}$ being an eigen-2-form of $\C_{\alpha\beta\lambda\mu}$
\be
\C_{\alpha\beta\mu\nu}\F^{\alpha\beta}\propto \F_{\mu\nu}\label{CF=F}
\ee
and this was the essential assumption in the several unique characterizations of the Kerr and other relative spacetimes given in references \cite{Mars1,Mars2,Mars3,Mars4,MS}.

The previous relation does not restrict the Petrov type of the space-time in general. However, a particular simple form where that condition is achieved is by assuming (recall (\ref{fsquare=0}))
\begin{align}
\C_{\alpha\beta\mu\nu} & = Q \left (\F_{\alpha\beta}
\F_{\mu\nu} - \frac{1}{3} \F^2 \I_{\alpha\beta\mu\nu} \right ) \label{C=FF1}  \\
& =Q \F_{\alpha\beta} \F_{\mu\nu} \label{C=FF}
\end{align}
which is actually the starting point in \cite{Mars1,Mars4,MS,MPSS,MPS}. We will assume this condition herein, even though we will make occasional comments and derive some results for the more general case above. If (\ref{C=FF}) holds, then at $p\in \M$ 
$$
k'^\alpha \C_{\alpha\beta\mu\nu} =0
$$
showing that the real null eigenvector of $\F_{\mu\nu}$ is the unique multiple principal null direction of the Weyl tensor so that the Petrov type is N if $Q|_p\neq 0$, or type 0 if $Q|_p=0$.

Under the assumption (\ref{C=FF})  equations 
(\ref{nablaF}) and (\ref{nablaFsq}) become, after using the definition
of Ernst one-form (\ref{defchi}), 
\begin{align}
\nabla_{\mu} \F_{\alpha\beta} &= \frac{1}{2} Q \chi_{\mu} \F_{\alpha\beta}
+ \frac{4}{3}  \Lambda \xi^{\nu} \I_{\nu\mu\alpha\beta},
\label{nablaF_2null} \\
\Lambda \chi_{\mu} & =0.  \label{cases}  
\end{align}
The vector field $\xi$ being Killing, it follows that $\pounds_{\xi}
\F_{\alpha\beta}=0$ and $\pounds_{\xi} \C_{\alpha\beta\mu\nu}=0$ and thus  
from (\ref{C=FF}) 
\be
\pounds_\xi Q =\xi^\mu \nabla_\mu Q =0 . \label{lieQ}
\ee
We want to elaborate the second Bianchi identity (\ref{divfree}) 
under the alignment condition. We start with the
following identity
\begin{align}
\F^{\alpha\beta} \xi^{\rho} \eta_{\rho\alpha\mu\nu} & = i \F^{\star}{}^{\alpha\beta} \xi^{\rho} \eta_{\rho\alpha\mu\nu}
= \frac{i}{2} \eta^{\alpha\beta\lambda\sigma} \F_{\lambda\sigma} \xi^{\rho} \eta_{\rho\alpha\mu\nu}  
 = \frac{i}{2} \delta^{\beta\lambda\sigma}_{\rho\mu\nu} \F_{\lambda\sigma} \xi^{\rho} 
\nonumber \\
& =
i \xi^{\beta} \F_{\mu\nu} + \frac{i}{2} ( \delta^{\beta}_{\nu} \chi_{\mu} - \delta^{\beta}_{\mu} \chi_{\nu} ) \label{Fxieta}
\end{align}
where in the last equality we have expanded $\delta^{\beta\lambda\sigma}_{\rho\mu\nu} 
= 3! \delta^{\beta}_{[\rho} \delta^{\lambda}_{\mu} \delta^{\sigma}_{\nu]}$ and have used the definition (\ref{defchi}).
We note that the identity (\ref{Fxieta})
is valid for any self-dual two-form and any vector field $\xi$, provided $\chi$ is defined as in (\ref{defchi}).
The second ingredient is the following identity which only holds for null two-forms as it relies on the expression
(\ref{nullF})
\begin{align}
\F_{\alpha\beta} \nabla_{\mu} \F^{\alpha}_{\phantom{\alpha}\nu} \nonumber
& = \F_{\alpha\beta} \left ( (\nabla_{\mu} k^{\prime \alpha}) m_{\nu}
+ k^{\prime \alpha} \nabla_{\mu} m_{\nu} 
- (\nabla_{\mu} m^{\alpha}) k^{\prime}_{\nu} - m^{\alpha} \nabla_{\mu} 
k^{\prime}_{\nu} \right ) \nonumber \\
& = (k^{\prime}_{\alpha} m_{\beta} - 
k^{\prime}_{\beta} m_{\alpha} )
\left ( (\nabla_{\mu} k^{\prime \alpha}) m_{\nu} - (\nabla_{\mu} 
m^{\alpha}) k^{\prime}_{\nu} \right ) \nonumber \\
& = (m_{\beta} k^{\prime}_{\nu} - k^{\prime}_{\beta} m_{\nu} ) 
m^{\alpha} \nabla_{\mu} k^{\prime}_{\alpha} \nonumber \\
& = -\F_{\beta\nu} m^{\alpha} \nabla_{\mu} k^{\prime}_{\alpha} 
\label{ident2}
\end{align}
where in the second equality we used that $k^{\prime}$ and $m$ lie in the
kernel of $\F$ and in the second one we used the orthogonality properties
of $k^{\prime}$ and $m$. The second Bianchi identity (\ref{divfree})
becomes 
\begin{align}
\nabla^{\alpha} \C_{\alpha\beta\mu\nu}
&= ( \nabla^{\alpha} Q \F_{\alpha\beta} - \Lambda Q \xi_{\beta} ) \F_{\mu\nu}
+ Q \F_{\alpha\beta} ( i \Lambda \xi^{\rho} \eta_{\rho\phantom{\alpha}\mu\nu}^{\phantom{\rho}\alpha} + \nabla_{\mu} \F^{\alpha}_{\phantom{\alpha}\nu}
- \nabla_{\nu} \F^{\alpha}_{\phantom{\alpha}\mu} ) \nonumber 
\\
& = ( \nabla^{\alpha} Q \F_{\alpha\beta} - 2 \Lambda Q \xi_{\beta} )
\F_{\mu\nu} 
+ 2 Q \F_{\beta[\mu} m^{\alpha} \nabla_{\nu]} k^{\prime}_{\alpha} =0
\label{Bianchi}
\end{align}
where in the first equality we used (\ref{maxwell}) and in the
second we inserted identities (\ref{Fxieta}) and (\ref{ident2})  and used the fact
that $\Lambda \chi_{\alpha}=0$, (cf. (\ref{cases})).

To conclude the section we find an identity for the derivative of the 
energy-momentum tensor $t_{\mu\nu}$ (\ref{emt}). Direct substitution of
(\ref{nablaF_2null})
yields
\begin{align*}
\nabla_{\mu} t_{\alpha\gamma} = \frac{1}{2} \nabla_{\mu} \left ( \F_{\alpha\beta}
\overline{\F}_{\gamma}^{\phantom{\gamma}\beta} \right ) =
\frac{1}{2} \left ( Q \chi_{\mu} + \overline{Q} \overline{\chi}_{\mu}
\right ) t_{\alpha\beta} + \frac{2}{3} \Lambda 
\xi^{\sigma} \underbrace{
\left ( \I_{\sigma\mu\alpha\beta} \overline{\F}_{\gamma}^{\phantom{\gamma}\beta}
+ \overline{\I}_{\sigma\mu\gamma\beta} \F_{\alpha}^{\phantom{\alpha}\beta} \right )}_{II}.
\end{align*}
To elaborate this further we need an expression for the second term $II$. We first
note the  identity
\begin{align*}
\X_{\alpha\beta} \overline{\Y}_{\gamma}^{\phantom{\gamma}\beta} 
- \X_{\gamma\beta} \overline{\Y}_{\alpha}^{\phantom{\alpha}\beta} = 0
\end{align*}
which is valid for any pair of self-dual two-forms $\bm{\X}$, $\bm{\Y}$
(this identity is a direct consequence of the even more fundamental
identity $X_{\alpha\beta} Y_{\gamma}^{\phantom{\alpha}\beta} - X^{\star}_{\gamma\beta}
Y_{\alpha}^{\star\beta} = \frac{1}{2} g_{\alpha\gamma} X_{\mu\beta}
Y^{\mu\beta}$ which holds for any two 2-forms \cite{Israel1970}). Applying this 
to $\I_{\sigma\nu\alpha\beta}$ (in the second pair of indices) and 
$\overline{\F}_{\gamma}^{\phantom{\gamma}\beta}$ gives
\begin{align}
II & = 
\I_{\sigma\mu\alpha\beta} \overline{\F}_{\gamma}^{\phantom{\gamma}\beta}
+ \overline{\I}_{\sigma\mu\gamma\beta} \F_{\alpha}^{\phantom{\alpha}\beta} =
\I_{\sigma\mu\gamma\beta} \overline{\F}_{\alpha}^{\phantom{\gamma}\beta}
+ \overline{\I}_{\sigma\mu\gamma\beta} \F_{\alpha}^{\phantom{\alpha}\beta} 
\nonumber \\
& = \frac{1}{4} \left ( g_{\sigma\gamma} ( \F_{\alpha\mu}
+ \overline{\F}_{\alpha\mu} )
- g_{\mu\gamma} ( \F_{\alpha\sigma} + \overline{\F}_{\alpha\sigma} ) \right )
+ \frac{i}{4} \eta_{\sigma\mu\gamma\beta} \left ( 
\overline{\F}_{\alpha}^{\phantom{\alpha}\beta}
- \F_{\alpha}^{\phantom{\alpha}\beta} \right )  \nonumber \\
& = \frac{1}{2} \left 
( g_{\sigma\gamma} F_{\alpha\mu} - g_{\mu\gamma} F_{\alpha\sigma}
- g_{\alpha\sigma} F_{\mu\gamma} - g_{\alpha\mu} F_{\gamma\sigma}
- g_{\alpha\gamma} F_{\sigma\mu} \right ) \label{ident3}
\end{align}
where in the third equality we inserted the explicit 
expression for $\I_{\sigma\mu\gamma\beta}$ and in the last one we 
used  $\overline{\F}_{\alpha}^{\phantom{\alpha}\beta}
- \F_{\alpha}^{\phantom{\alpha}\beta} = - 2 i F_{\alpha}^{\star\phantom{\alpha}\beta}
= - i \eta_{\alpha\phantom{\beta}\rho\kappa}^{\phantom{\alpha}\beta} F^{\rho\kappa}$
and have expanded the product of volume forms, as in (\ref{Fxieta}).
Identity (\ref{ident3}) is valid for any self-dual two-form $\F_{\alpha\beta}
= F_{\alpha\beta} + i F^{\star}_{\alpha\beta}$. Contracting $II$ with $\xi^{\sigma}$ and
using that $\Lambda \chi_{\mu} =0$ we finally arrive at
\begin{align}
\nabla_{\mu} t_{\alpha\gamma} 
= \frac{1}{2} \left ( Q \chi_{\mu} + \overline{Q} \overline{\chi}_\mu \right )
t_{\alpha\gamma} + \frac{\Lambda}{3} \left ( \xi_{\gamma} F_{\alpha\mu}
+ \xi_{\alpha} F_{\gamma\mu} \right )
\label{nablatnull}
\end{align}
It follows from e.g. (\ref{cases}) that 
the cases $\Lambda=0$ and $\Lambda \neq 0$ are different from
each other, and thus we treat them separately.


\section{The case $\Lambda=0$}
\label{Lambda0}
From Lemma \ref{lem:F=0} we know that the self-dual Killing 2-form $\F$ can vanish on an open neighbourhood only if $\Lambda =0$. Thus, we first consider this possibility now. In this case, the main assumption (\ref{C=FF}) implies that the Weyl tensor vanishes, so the spacetime has a vanishing Riemann tensor and is locally flat. The alignment condition
(\ref{C=FF}) is satisfied by any of the ten linearly independent Killing vectors
(by simply choosing $Q=0$)
and the subclass for which $\F=0$ is the four-dimensional
space of covariantly constant vector fields (i.e.
the translations). This is rather an exceptional situation and, as we will see, is included naturally as a limit in the solutions of the generic case with $\F\neq 0$, which we consider next.

The next proposition goes a long way in identifying the spacetimes
contained in this class.
\begin{Proposition}
\label{PropLambda0}
Let $(\M,g)$ be a vacuum spacetime with a Killing
vector $\xi$ such that
the self-dual Killing form $\F_{\alpha\beta}$ is null and nowhere zero and
the self-dual Weyl tensor takes the form (\ref{C=FF}).
Then $(\M,g)$ admits a smooth, null and parallel local vector
field $k$. Moreover, $\xi$ and $k$ commute and $g(\xi,k)$ is constant. In the case that this constant is non-zero, $k$ is defined globally and can be chosen so that $g (\xi, k) = - 1$ 
(in particular, $\xi$ vanishes nowhere on $\M$).
\end{Proposition}

\vs 

\proof
From (\ref{emt}) and (\ref{nablatnull}) with $\Lambda =0$ we have
$$
2\nabla_\lambda k'_{(\mu} k'_{\nu )}=\frac{1}{2}(Q\chi_\lambda +\overline Q \overline\chi_\lambda)k'_\mu k'_\nu
$$
which is equivalent to 
\be
\nabla_\lambda k'_\mu =\frac{1}{4}(Q\chi_\lambda +\overline Q \overline\chi_\lambda)k'_\mu \label{nablak'}
\ee
meaning that the vector field $k'$ is recurrent, and in particular $m^\mu \nabla_\lambda k'_\mu =0$ so that the Bianchi identity (\ref{Bianchi}) 
simplifies to $\F_{\alpha\beta}\nabla^\alpha Q=0$. Thus
$\nabla_\alpha Q = C k'_\alpha +B m_\alpha$
which together with (\ref{lieQ}) implies
\be
C\xi^\alpha k'_\alpha +B \xi^\alpha m_\alpha =0 . \label{step}
\ee
A (local) vector field in the direction of $k'$ will then be parallel if the proportionality one-form $Q\chi_\lambda +\overline Q \overline\chi_\lambda$ is closed. This is the case as follows from the following calculation. Using that 
$\chi$ is closed (\ref{chiclosed}),
$$
d(Q \chi) =dQ\wedge \chi = 2(C k' + B m)\wedge [(\xi^\alpha k'_\alpha)m - 
(\xi^\alpha m_\alpha) k']=
2(C\xi^\alpha k'_\alpha +B \xi^\alpha m_\alpha )k' \wedge m =0
$$
where we have used (\ref{chinull}) in the second equality
and (\ref{step}) in the last one. It follows that locally there exists a real function $w$ (defined
up to an additive constant)
such that $Q\chi_\lambda +\overline Q \overline\chi_\lambda =4 \nabla_\lambda w$ 
and (\ref{nablak'}) reads
$$
\nabla_\lambda \left(e^{-w}k'_\mu \right) =0
$$
so that $k\defi e^{-w} k'$ is a covariantly constant null vector field. Taking into account that $\xi^\mu\nabla_\mu w=0$ and (\ref{liek'}), 
\be
\pounds_\xi k =0 \label{liek}
\ee
so that $\xi$ and $k$ commute. Moreover the scalar product $\xi^\rho k_\rho$ is constant because
\begin{align*}
\nabla_{\alpha} \left ( \xi_{\rho} k^{\rho} \right ) = 
\left ( \nabla_{\alpha} \xi_{\rho} \right ) k^{\rho}
= \mbox{Re} \left ( \F_{\alpha\rho} \right ) k^{\rho} =0
\end{align*}
because $k$ lies in the kernel of $\F$.
Thus $g(\xi,k)$ is either zero everywhere on $\M$ or else vanishes 
nowhere. 

In the latter case we can choose the 
additive constant in $w$ (i.e. scale $k$ by a constant)
so that $g(\xi,k)=-1$ everywhere. It follows that $k$ is defined not only locally
on $\M$ but also globally. To see this, assume that $g(\xi,k') \neq 0$
somewhere. Let $\M^{\xi}$ be the set of points where $\xi$ is not zero.
Given that $k'$ is smooth, nowhere zero and
globally defined on $\M$,
a smooth vector field $k$ (we use the same notation because
{\em a posteriori} this vector field is the same
as before) 
can be uniquely defined on
$\M^{\xi}$ by the conditions of being proportional to $k'$ and 
satisfying $g (\xi,k)=-1$. The property of being recurrent is invariant
under scaling, so $k$ is also recurrent and we have
$\nabla_{\alpha} k_{\beta} = W_{\alpha} k_{\beta}$ for some 
smooth one-form $W_{\alpha}$. But then
\begin{align*}
0 = \nabla_{\alpha} \left ( \xi_{\rho} k^{\rho} \right )
= \mbox{Re} \left ( \F_{\alpha\rho} \right ) k^{\rho}
+ \xi^{\rho} W_{\alpha} k_{\rho} = - W_{\alpha} 
\end{align*}
where again we used $\F_{\alpha\rho} k^{\rho} =0$. Thus
$k$ is in fact parallel and smooth on $\M^{\xi}$. It must extend
smoothly to the closure $\overline{\M^{\xi}}$ (consider e.g. 
geodesics $\gamma$ starting at $p \in \partial \M^\xi$ into $\M^{\xi}$ and
use the property that $g(\dot{\gamma},k)$ is constant along the geodesic) and
hence to all of $\M$. Since $g(\xi,k)$ also extends 
smoothly to all of $\M$ (and takes the value $-1$), we in fact conclude that
$\xi$ had no
fixed points on $\M$ after all. \hfill $\Box$

\vspace{5mm}

We can now find explicitly the spacetimes satisfying 
the hypotheses of Proposition \ref{PropLambda0}.
Spacetimes with a covariantly constant null vector field $k$ are called
Brinkmann spaces \cite{Bri}. Schimming \cite{Schimming} has shown that Brinkmann spaces admit local coordinates $\{v,u,x,y \}$ near any point $p \in  \M$ such that 
\be
k =\partial_v, \hspace{1cm} \bm{k} =-du \label{kparallel}
\ee
and the metric takes the form
\begin{align}
ds^2 = - 2 dv du + 2H du^2 + dx^2 + dy^2,
\label{localform}
\end{align}
where $H(u,x,y)$ is  a function not dependent on $v$ (as $k$ is, in particular, a Killing vector). These spacetimes are vacuum if and only if $H$ solves the elliptic PDE 
\be
H_{,xx} + H_{,yy} =0. \label{LaplaceH}
\ee 
The local coordinates are not fully fixed and the form (\ref{localform}) together with (\ref{kparallel}) are kept under several coordinate changes, see e.g. \cite{Exact,BSS}, in particular under the local transformation
\be
\label{coord}
u=u', \, \, x=x'+p_1(u), \, \,  y=y'+p_2(u), \, \,  v=v'+\dot{p}_1 x' +\dot{p}_2 y' +s(u)
\ee
where $p_1(u)$ and $p_2(u)$ are functions of only $u$, dots stand for derivative with respect to $u$, and the new function $H'(u',x',y')$ is
$$
H' = H +\ddot{p}_1 x' +\ddot{p}_2 y' +\dot{s}+\frac{1}{2}  \left(\dot{p}_1^2 +\dot{p}_2^2 \right).
$$
It follows that any terms linear on $x$ and $y$ (with $u$-dependent coefficients) appearing in $H$ can be removed by such a transformation, while keeping the vacuum condition (\ref{LaplaceH}). We will use this freedom presently.

We want to identify which subclass of Brinkmann spacetimes 
are included in the spacetime satisfying the hypotheses of Proposition 
\ref{PropLambda0}. Our strategy is to identify the null self-dual two-forms
$\F_{\alpha\beta}$ associated to a Killing vector $\xi$ in the coordinates where (\ref{localform}) holds. 
Then, we will also require that the alignment condition (\ref{C=FF}) holds.

To that end,  observe that the metric (\ref{localform}) takes a Kerr-Schild form \cite{Exact}
\be
g= \eta +2 H \bm{k}\otimes \bm{k} \label{KS}
\ee
where $\eta$ is the flat Minkowski metric. Hence, any Killing vector $\xi$ of the metric with the property (\ref{liek}) satisfies
$$
\pounds_\xi g =\pounds_\xi \eta +2 \xi(H)\, \bm{k}\otimes \bm{k}=0 \hspace{1cm}\Longrightarrow  \hspace{1cm} \pounds_\xi \eta =-2 \xi(H)\,  \bm{k}\otimes \bm{k} 
$$
and these are called Kerr-Schild vector fields \cite{CHS} of the Minkowski spacetime relative to the parallel null direction $k$. The general solution for such Kerr-Schild vector fields was found in \cite{CHS} and, taking into account (\ref{liek}) they read (adapting the notation)
$$
\xi = - g(\xi,k) \partial_u 
+ \left ( \dot{a} x + \dot{b} y + c (u)\right ) \partial_v
+ (a(u) +\alpha y) \partial_x + (b(u) -\alpha x) \partial_y
$$
for arbitrary real functions $a(u),b(u),c(u)$ and constant $\alpha\in \mathbb{R}$, and with
\be
\xi (H) = \dot c +\ddot{a} x +\ddot{b} y \label{Heq} .
\ee
To check which of these vector fields produce a null Killing 2-form, we first note that
$$
\bm{\xi} = g(\xi,k) dv -\left ( \dot{a} x + \dot{b} y + c (u)+2H g(\xi,k) \right )du + \left(a(u) +\alpha y\right) dx + (b(u) -\alpha x) dy
$$
so that
$$
d\bm{\xi} = 2du\wedge \left[\left(\dot a +g(\xi,k) H_{,x} \right) dx +\left(\dot b +g(\xi,k) H_{,y} \right)dy\right]+2\alpha dy\wedge dx
$$
and the 2-form $d\bm{\xi}$ is null, with $k$ as null eigenvector, if and only if $\alpha =0$, which we assume henceforth. Therefore,
\begin{align}
\xi = - g(\xi,k) \partial_u 
+ \left ( \dot{a} x + \dot{b} y + c(u) \right ) \partial_v
+ a(u) \partial_x + b(u) \partial_y\, ,
\label{Killing}
\end{align}
the PDE (\ref{Heq}) becomes
\be
\xi (H) = -  g(\xi,k) H_{,u }
+  a(u) H_{,x} + b(u) H_{,y} = \dot c +\ddot{a} x +\ddot{b} y \label{pde}
\ee
and the Killing null 2-form $F=d\bm{\xi}/2$ and its dual are (the basis $\{ \partial_v,
\partial_u,\partial_x,\partial_{y}\}$ is taken as positively oriented)
\bean
F = du\wedge \left[\left(\dot a +g(\xi,k) H_{,x} \right) dx +\left(\dot b +g(\xi,k) H_{,y} \right)dy\right],\\
F^\star = du\wedge \left[-\left(\dot b +g(\xi,k) H_{,y} \right) dx +\left(\dot a +g(\xi,k) H_{,x} \right) dy\right]. 
\eean
Thus, we have
\be
\F = \left(\dot a -i \dot b +g(\xi,k) (H_{,x} -i H_{,y}) \right) du \wedge (dx+i dy) \label{selfF} .
\ee
Given the simple form of the metric (\ref{localform}), it is a matter of
straightforward calculation to compute the self-dual Weyl tensor $\C_{\alpha\beta\mu\nu}$, which turns out to be
\begin{align}
\C_{\alpha\beta\mu\nu} = - \frac{1}{2} \left ( 
H_{,xx} - H_{,yy} -  2 i H_{,xy} \right ) \V_{\alpha\beta}  \V_{\mu\nu}
\label{align2}
\end{align}
where $\V$ is the self-dual two form 
\begin{align*}
\V = \bm{k} \wedge (dx + i dy) =-du \wedge (dx + i dy) .
\end{align*}
Thus, our main assumption, the alignment condition (\ref{C=FF}), is clearly satisfied for (\ref{Killing}) as follows from (\ref{align2}) and (\ref{selfF}).

Now it becomes necessary to split the analysis into two cases,  namely
when $g(\xi,k) = -1$ and when $g(\xi,k)=0$.

In the case $g(\xi,k)=0$, 
equation (\ref{pde}) implies after derivation 
with respect to $x$, $y$ and taking 
(\ref{LaplaceH}) into account 
$$
a(u) H_{,xx} +b(u) H_{,xy} =\ddot{a}, \hspace{1cm} a(u) H_{,xy} -b(u) H_{,xx}= \ddot{b}
$$
so that $H_{,xx}$ and $H_{,xy}$ depend only on $u$ 
(note that $a(u)$
and $b(u)$ cannot vanish simultaneously on an open interval because
$\F$ is assumed to be non-zero, cf.\ (\ref{selfF}) with $g(\xi,k)=0$).
It follows that
there exist five (real) functions $A(u), B(u), s_1(u), s_2(u), s_3(u)$
such that
\begin{align}
H = \frac{1}{2} A (u) \left ( x^2 - y^2 \right ) +B(u) x y + s_1(u) x +s_2 (u) y + s_3(u)
\label{formh} 
\end{align}
and the equation (\ref{pde}) requires
\begin{align}
\ddot{a} & = A a + B b,  \quad  \quad   \ddot{b} = B a - A b, \label{system}\\ 
\dot c & =s_1 a +s_2 b .  \label{system2} 
\end{align}
However, as mentioned after equation (\ref{coord}), all linear terms in $x$ and $y$ with arbitrary $u$-dependent coefficients can be removed in (\ref{formh}) by means of the transformation (\ref{coord}), and thus we can assume, without loss of generality, that $s_1=s_2=s_3=0$ which in particular implies, from (\ref{system2}), that $c=c_0$ is a constant.\footnote{The explicit coordinate change (\ref{coord}) that achieves this has $\{ p_1,p_2,s\}$ solving the system of ODEs
\begin{align*}
\ddot{p}_1 &= A p_1 + B p_2 +  s_1, \quad \quad
\ddot{p}_2 = B p_1 -A p_2 + s_2, \\
2 \dot{s} &=  2s_3 + A (p_1^2 - p_2^2) + 2 B p_1 p_2
+ 2s_1 p_1 + 2s_2 p_2 +\dot{p}_1^2 + \dot{p}_2^2.
\end{align*}
}
The general solution of the system (\ref{system}) has four integration
constants. We have thus found that the most general local form
of a spacetime satisfying the
hypotheses of Proposition \ref{PropLambda0} with $g(\xi,k)=0$
can be written as 
\begin{align}
ds^2 & = - 2 du dv + dx^2 + dy^2 + \left (A(u) (x^2 - y^2) + 2 B(u) x y \right)
du^2, \label{firstCase} \\
\xi & = ( \dot{a} x + \dot{b} y + c_0 ) \partial_v +  a \partial_x
+ b \partial_y, \label{firstCaseKill}
\end{align}
where $c_0 \in \mathbb{R}$ and $\{ a(u), b(u)\}$  is any solution of (\ref{system}). If $A(u) = B(u) =0$ on some non-empty set $\U\subset \M$ then
$(\U,g)$ is obviously locally flat leading to the exceptional case mentioned at the beginning of this section.
If, on the other hand, $A(u)$  and $B(u)$ do not
vanish simultaneously on any open set $\U'$, these metrics are called vacuum {\em plane waves}
\cite{BJ,Exact}, they admit the 5 parameter family of Killing vectors
(\ref{firstCaseKill}).  
The alignment condition (\ref{C=FF}) holds for {\em any} of these Killing
vectors. The Killing form is (see (\ref{selfF}))
\begin{align*}
\F_{\alpha\beta} = ( - \dot{a} + i \dot{b}) \V_{\alpha\beta}
\end{align*}
and the proportionality function $Q$ is, from (\ref{align2}),
\begin{align*}
Q 
= - \frac{A - i B}{(\dot{a} - i \dot{b} )^2}.
\end{align*}
The set of points where $\F_{\alpha\beta} =0$ (i.e. those 
with $\dot{a} = \dot{b} =0$), which were excluded by
the hypothesis of $\F$ being non-zero, can be attached to the spacetime
at the cost of $Q$ being non-smooth there. However, this is merely an artifact of
the proportionality function between the Killing form $\F_{\alpha\beta}$ and
the two-form $\V_{\alpha\beta}$  becoming zero.
The extended spacetime obviously remains smooth also at those points.

The case $g(\xi,k)=-1$ is simpler. Consider the coordinate transformation
(\ref{coord}) with $p_1,p_2,s$ satisfying
\begin{align*}
  \dot{p}_1 = a, \quad \quad \dot{p}_2 = b, \quad \quad
\frac{d}{du} \left ( s- p_1 \dot{p}_1- p_2 \dot{p}_2 \right )
= \dot{c} - a \dot{p}_1 - b \dot{p}_2.
\end{align*}
The Killing vector $\xi$ in (\ref{Killing})
is simply $\xi = \partial_{u}$ in the new coordinates.
It follows that we can
assume without loss
of generality $\xi= \partial_u$ (i.e. $a=b=c=0$) in the expressions above.
The PDE (\ref{pde}) states simply that $H(x,y)$ is
independent of $u$. Therefore, these are precisely the {\em stationary} vacuum Brinkmann spacetimes (or the stationary vacuum pp-waves) considered in \cite{H}
\begin{align}
ds^2 = - 2 du dv + dx^2 + dy^2 + 2H(x,y) du^2, \quad \quad H_{,xx} + H_{,yy}=0,
\quad \quad \xi= \partial_u.
\label{Stationary}
\end{align} 
This metric admits in general a 2-parameter family of Killing vectors generated by the parallel $k$ and by $\xi$. The length of the 2-form $\bm{k} \wedge \bm{\xi}$ is constant.

Note that $H$ being a solution of the Laplace equation (\ref{LaplaceH}), $H$
is the real part of some complex function $\sigma$ holomorphic
in the variable $\zeta = x + i y$. We also have
$$
\F_{\alpha\beta} = (H_{,x}-iH_{,y}) \V_{\alpha\beta} =\sigma_{,\zeta} \V_{\alpha\beta}.
$$ 
The alignment condition (\ref{align2}) can be rewritten as
\begin{align*}
\C_{\alpha\beta\mu\nu} = - \sigma_{,\zeta\zeta} \V_{\alpha\beta}
\V_{\mu\nu}
\end{align*}
so that, since we are assuming $\F \neq 0$, the function $Q$ reads 
\begin{align*}
Q   
=  - \frac{ \sigma_{,\zeta\zeta}}{(\sigma_{,\zeta})^{2}}.
\end{align*}
As in the previous case, the set of points where $\F_{\alpha\beta} =0$, which are given by the zeros of the holomorphic function $\sigma_{\zeta}$ and were excluded by assumption, can be attached to the spacetime at the cost of making $Q$ non-smooth there, but keeping the extended spacetime fully smooth also at those points.

\vspace{2mm}
{\bf Remark:}
It must be noticed that there is a non-empty intersection of the two families of spacetimes identified in this section. They are given by the analytic function $\sigma = \beta\zeta^2$ with $\beta\in \mathbb{C}$,
$2 \beta = A-iB$ ---where now $A,B$ are constants. 
In this case there is a 6-parameter family of Killing vectors satisfying the alignment (\ref{C=FF}). These metrics are precisely the {\em irreducible} locally symmetric vacuum spacetimes, that is, they satisfy the condition $\nabla_\rho R_{\alpha\beta\mu\nu}=0$, and were identified in \cite{CW}, see also
\cite{BSS}. 
For later reference, we write down explicitly the metric and the Killing
vectors. Define constants  $\kappa>0$ and $\alpha$ by
$A = \kappa^2 \cos (2 \alpha)$ and $B = \kappa^2 \sin (2 \alpha)$. Then,
the coordinate change $x' = x \cos \alpha  + y \sin \alpha $,
$y' = - x \sin \alpha  + y \cos \alpha$  transforms the metric (after
dropping the primes)  
into (\ref{firstCase}) with $A = \kappa^2>0$ and $B=0$. 
The solution of
(\ref{system}) (with $A = \kappa^2, B=0$) is 
\begin{align}
a(u)  = c_1 e^{\kappa u} + c_2 e^{-\kappa u}, \quad \quad
b(u) = c_3 \cos ( \kappa u ) + c_4 \sin (\kappa u ),
\label{ab}
\end{align}
where $c_1,c_2,c_3,c_4$ are arbitrary constants.  Thus, the metric
and Killing vectors are
\begin{align}
ds^2  = & -2 dudv + dx^2 + dy^2 + \kappa^2 (x^2 - y^2) du^2, \label{LocSym1}\\
\xi  = & \left (c_0 + 
\kappa x ( c_1 e^{\kappa u} - c_2 e^{-\kappa u})  
+ \kappa y (-c_3 \sin(\kappa u) + c_4 \cos(\kappa u) )   \right )
\partial_v + \nonumber \\
& + (c_1 e^{\kappa u} + c_2 e^{- \kappa u} ) \partial_x
+ (c_3 \cos(\kappa u) + c_4 \sin(\kappa u)) \partial_y 
+ c_5 \partial_u. \label{LocSym2}
\end{align}
The alignment condition is satisfied by any non-trivial Killing vector
in (\ref{LocSym2}). If $c_5 \neq 0$ we are in the case 
$g(\xi,k) \neq 0$ while $c_5=0$ corresponds to $g(\xi,k)=0$. The proportionality factor
$Q$ in (\ref{C=FF}) for the general Killing (\ref{LocSym2}) takes the form
(with $a(u),b(u)$ as in (\ref{ab}))
\begin{align*}
Q = \frac{-\kappa^2}{\left ( \dot{a} - i \dot{b} - c_5 \kappa^2 (x + i y) 
\right )^2}.
\end{align*}

\section{The case $\Lambda\neq 0$}
\label{Lambdaneq0}
\vs In the $\Lambda$-vacuum case we start with the following result
\begin{Proposition}
\label{Prop2}
Let $(\M,g)$ be a $\Lambda$-vacuum ($\Lambda \neq 0$) 
spacetime with a Killing
vector $\xi$ such that
the self-dual Killing form $\F_{\alpha\beta}$ is null 
and such that the self-dual Weyl tensor takes the form (\ref{C=FF}).
Then 
$\Lambda <0 $, both $\F_{\alpha\beta}$ and $\xi$ vanish nowhere and  $\xi$ 
is null, hypersurface orthogonal, pointing along the principal null direction of $\F_{\alpha\beta}$.
Moreover, the quotient metric on each null hypersurface
orthogonal to $\xi$ is of negative constant curvature $R^{(2)} =  \frac{2 \Lambda}{3}$.
\end{Proposition}

\proof  
Consider the set $\U$ where $\F_{\alpha\beta}$ is not zero. To start with, $\U$ cannot be empty, as otherwise $\F$ would vanish on $\M$ which is impossible for $\Lambda\neq 0$ according to Lemma \ref{lem:F=0}.
From (\ref{cases}), on $\U$  we have $\chi_\mu =0 =\xi^\nu \F_{\nu\mu}$, so that $\xi$ must be proportional to the unique principal null direction $k $ of $\F_{\mu\nu}$. 
Thus 
\be
\xi_\alpha = L k'_\alpha  \label{xi=Ak}
\ee
for some function $L$. Note that $\xi$ is automatically null, so $N=0$.
$L$ is in fact  a constant directly related to
$\Lambda$ as follows  from (\ref{derchi}) with $\chi_\mu =0$ and $N=0$
$$
0=-k'_\mu k'_\nu -\frac{2}{3}\Lambda \xi_\mu \xi_\nu =-k'_\mu k'_\nu \left(1+\frac{2}{3}\Lambda L^2 \right)
$$
leading to
\be
\Lambda =-\frac{3}{2L^2} <0 \, .
\ee
From (\ref{emt}) we have $2 t_{\mu\nu} = 2 \F_{\mu\rho} \overline{\F}_{\nu}^{\phantom{\nu}\mu} = - \frac{2 \Lambda}{3} \xi_{\mu}\xi_{\nu}$. This shows that 
$\partial \U$ is empty as otherwise, at 
$p \in \partial \U$, we would have
both $\F_{\alpha\beta}|_p=0$ and $\xi_{\alpha}|_p=0$
which cannot happen for non-identically zero Killing vectors. Thus, $\U = \M$
and neither $\F_{\alpha\beta}$ nor $\xi$ vanish anywhere.

Equation (\ref{nablatnull}) becomes, on using (\ref{emt}),
$$
k'_\gamma \left(\nabla_\mu k'_\alpha -\frac{2 \Lambda}{3} L F_{\alpha\mu} \right)+
k'_\alpha \left(\nabla_\mu k'_\gamma -\frac{2 \Lambda}{3} L F_{\gamma\mu} \right)=0
$$
so that
\be
\nabla_\mu k'_\alpha =\frac{2\Lambda}{3} L F_{\alpha\mu}
= - \frac{1}{2L} 
\left[ k'_\alpha (m_\mu +\overline m_\mu) - k'_\mu (m_\alpha +\overline m_\alpha)\right] 
\label{nablaknull} .
\ee
This expression implies, in particular, that $k'_{\alpha}$ (and hence $\xi_{\alpha}$)
is hypersurface orthogonal. 
Thus, by the Fr\"obenius theorem, through 
each point $p \in \M$ there passes a unique, maximal, injectively immersed
null hypersurface $\H_{p}$ orthogonal to $\xi$. We call such 
hypersurfaces ``horizons''.
From (\ref{nablaknull}) we know that any spacelike section $S$ of $\H_{p}$ has a vanishing second fundamental form along $k'$, which also implies that the mean curvature vector of $S$ in $\M$ is null. This, together with the fact that $k'^\mu C_{\mu\nu\alpha\beta}=0$ and using the standard form of the Gauss equation for $S$ (see e.g. formula (9) in \cite{S2}, the formula previous to (7) in \cite{Mars5}, or eq.(16) in \cite{HM}) leads to
\begin{eqnarray*}
K_S =\frac{R^{(2)}}{2}=\frac{\Lambda}{3} 
\end{eqnarray*}
where $K_S$ represents the Gaussian curvature of $S$ and and $R^{(2)}$ its scalar curvature. \hfill $\Box$

\vspace{5mm}

We want to use the information provided in the proof of this Proposition
to determine the local form of the spacetime metric.
To that aim we use that $\bm{k'}$ is integrable and nowhere zero
and introduce two smooth functions $f$ and $u$ such that 
\begin{align}
\bm{k}^{\prime} = -f du.
\label{kk}
\end{align}
Since neither $f$ nor $du$ vanish anywhere we can, without loss
of generality, assume that $f > 0$. Let us find the PDE's that this function satisfies. 
Combining the exterior differential of (\ref{kk}) $d \bm{k}^{\prime} = f^{-1} df \wedge \bm{k}^{\prime}$  with expression (\ref{nablaknull}) 
yields
\begin{align}
\frac{1}{L} \left ( m_{\alpha} + \overline{m}_{\alpha} \right ) = 
 - \frac{1}{f} \nabla_{\alpha} f  + h_1 k'_{\alpha}
\label{ss}
\end{align}
for some smooth function $h_1$. Note that this expression
implies in particular that 
\begin{align}
\nabla_{\alpha} f \nabla^{\alpha}f = \frac{2 f^2}{L^2} .
\label{normsq}
\end{align}
Inserting (\ref{kk}) and (\ref{ss}) into (\ref{nablaknull}) it follows
\begin{align*}
\nabla_{\mu} \nabla_{\alpha} u = - \frac{1}{2f} \left ( \nabla_{\alpha} u
\nabla_{\mu} f + \nabla_{\mu} u \nabla_{\alpha} f \right ).
\end{align*}
We can evaluate the integrability conditions of this equation. Using the
fact that
\begin{align*}
C^{\rho} _{\phantom{\rho}\alpha\mu\beta} \nabla_{\rho} u
= - \frac{1}{f} C^{\rho} _{\phantom{\rho}\alpha\mu\beta} k'_{\rho} 
= - \frac{1}{f} \mbox{Re} \left ( \C^{\rho} _{\phantom{\rho}\alpha\mu\beta} k'_{\rho} \right ) 
=0,
\end{align*}
the integrability condition turns out to be
\begin{align*}
\nabla_{[\beta} u \left (  
\frac{1}{f} \nabla_{\mu]} \nabla_{\alpha} f
- \frac{1}{2 f^2} \nabla_{\mu]} f \nabla_{\alpha} f  - \frac{1}{L^2} g_{\mu]\alpha}
\right ) = 0
\end{align*}
which is equivalent to 
\begin{align}
\nabla_{\mu} \nabla_{\alpha} f
- \frac{1}{2 f} \nabla_{\mu} f \nabla_{\alpha} f  - \frac{1}{L^2} f g_{\mu\alpha}
= h_2 \nabla_{\mu} u \nabla_{\alpha} u 
\label{Hess0}
\end{align}
for some smooth function $h_2$.   At this point, it is convenient
to define a smooth positive function $x$ by $f:=  \frac{2 L}{x^2}$.
In terms of $x$, the norm condition (\ref{normsq}) and the Hessian equation (\ref{Hess0})
become
\begin{align}
\nabla_{\alpha} x \nabla^{\alpha}x & = \frac{x^2}{2L^2}, \label{norm}\\
\nabla_{\mu} \nabla_{\alpha} x & - \frac{2}{x} \nabla_{\mu} x \nabla_{\alpha} x
+ \frac{x}{2L^2} g_{\mu\alpha} = - \frac{h_2 x^3}{4 L} \nabla_{\mu} u
\nabla_{\alpha} u\label{Hess}
\end{align}
and we also have, from (\ref{ss}),
\be
k'^{\alpha}\nabla_{\alpha} x =0. \label{kx}
\ee
We now construct a local coordinate system near any point $p \in \M$.
Consider a hypersurface $\Sigma$ passing through $p$
and transversal to the horizon $\H_{u(p)} := \{u = u(p) \}$. Consider an
open neighbourhood $U_p$ of $p$ and define $I_u \subset \mathbb{R}$
as the set of values of $u$ such that the horizons 
$\H_{u}$ of constant $u$ intersect $U_p$. Restricting $U_p$ if necessary
we can fulfill the following three conditions:
\begin{itemize}
\item $I_u$ is connected, 
\item $\Sigma$ is transverse
to  $\H_u$ for all $u \in I_u$,
\item the spacelike surface
$S_{u} = \Sigma \cap \H_{u}$ is non-empty and connected for all $u
\in I_u$. 
\end{itemize}
Let $S_{u,v}$ be the set for surfaces obtained by Lie
dragging $S_u$ along the (affinely parametrized) null generator
with tangent $\xi$ of $\H_{u}$ and satisfying $S_{u,v=0} = S_{u}$.
Restricting $U_p$ further we can assume that $v$ takes values on an
open interval $I_v$ containing zero and that all $S_{u,v}$, $\forall v \in I_v$
are diffeomorphic to $S_u$. Moreover we can assume all $S_u$, $u \in I_u$
to be simply connected. 

Consider the function $x_S:= x|_{S_{u,v}}$ restricted
to the surface $S_{u,v}$ and endow $S_{u,v}$ with the induced metric
$h$ and the corresponding Levi-Civita covariant derivative, which we denote by
$D$. The pull-back of the Hessian of a function $F$
and the Hessian of the restriction $F_\mathcal{S}$ on an embedded
submanifold $\mathcal{S}$ with embedding  $\varphi$ are related by
\begin{align}
\mbox{Hess}^D \, F_\mathcal{S} = \varphi^{\star} (\mbox{Hess} \, F) 
- g(\ii, \grad \, F) \label{idenHess}
\end{align}
where $\ii$ is the shape tensor ---also called second fundamental form vector --- of $\mathcal{S}$ in $\M$ (see e.g. \cite{O}).
In order to apply this to $x_S$ we observe
that the shape tensor of $S_{u,v}$ satisfies $g(\ii,k^{\prime} ) =0$, because
$S_{u,v}$ lies in a horizon, which as seen in the proof of Proposition \ref{Prop2} are null hypersurfaces with identically vanishing second 
fundamental form. This means that $\ii$ is parallel to $k^{\prime}$
and we have  $g( \ii, \grad \, x )=0$ as a consequence of (\ref{kx}). 
Pulling-back (\ref{Hess}) onto $S_{u,v}$ by means of (\ref{idenHess}) we thus find
\begin{align}
D_A D_B x_S - \frac{2}{x_S} (D_A x_S) (D_B x_S) + \frac{x_S}{2L^2} h_{AB} =0
\label{Hess2}
\end{align}
where $h_{AB}$ is the induced metric on $S_{u,v}$.
Moreover, from (\ref{norm}) and (\ref{kx})
the square norm of $x_S$ satisfies
$$
(D_A x_S) (D^A  x_S) = \frac{x_S^2}{2 L^2}.
$$
 The trace of (\ref{Hess2}) is therefore 
$$
D_A D^A x_S =0
$$ 
so that $x_S$ is a harmonic function with $dx_S$ nowhere vanishing. 
Its Hodge dual $\star dx_S$ in $S_{u,v}$ is then a closed
one-form and simply connectedness of $S_{u,v}$ implies the existence
of a function $y_S$ satisfying $\star dx_S = dy_S$. By construction $dx_S$ and 
$dy_S$ are mutually orthogonal and have the same norm, which implies that the metric
$h_{AB}$ takes the form
\begin{align*}
h = \frac{2L^2}{x_S^2} \left ( dx_S^2 +  dy_S^2 \right ).
\end{align*}
This metric has constant negative curvature equal to
$- 1/(2L^2)$ in accordance with Proposition \ref{Prop2}. The pair
$\{x_S,y_S\}$ is a coordinate system of $S_{u,v}$. The function $y_S$
is defined up to an additive constant on each $S_{u,v}$. We can fix
partially this constant by selecting, on each $\H_u$ one
null generator $\gamma(v)$ and imposing $y_S |_{\gamma(v)} = y_0$ 
for some fixed constant $y_0$. 

By construction
the set of coordinates which assigns to each point $q \in U_p$
the values $\{u,v\}$ of the surface $S_{u,v}$ containing  $q$ and
the values $\{x_S,y_S\}$ of its coordinates within $S_{u,v}$ is a
coordinate system of $U_p$. We will denote this coordinate system by
$\{v,u,x,y\}$ (we can use $x$ because by construction this coordinate
agrees with the function $x$ introduced before). We already know
that $k^{\prime \alpha} \nabla_{\alpha} u =0$ from (\ref{kk}), and we also have (\ref{kx}) 
and $k^{\prime \alpha} \nabla_{\alpha} v = 1/L$, the latter as a consequence of $v$ 
being the affine parameter associated to $\xi = L k^{\prime}$.
Moreover $k^{\prime \alpha} \nabla_{\alpha} y =0$ , as we show next. 
We work within a fixed $\H_{u_0}$. Let $W_1,W_2$ be vectors tangent to $S_{u_0,v}$ satisfying $[k^{\prime},W_1]
=[k^{\prime},W_2]=0$. $k'$ being an isometry implies that
\begin{align*}
0 & = \pounds_{k'} (g( W_1,W_2) )= \pounds_{k'} (h(W_1,W_2)) = 
\pounds_{k'} \left ( \frac{2L^2}{x^2} \left ( dx(W_1) dx(W_2) +  
dy(W_1) dy(W_2) \right ) \right ) \\
& = \frac{2L^2}{x^2}  d (\pounds_{k^{\prime}} y) (W_1) 
d (\pounds_{k^{\prime}} y) (W_2) 
\end{align*}
where in the last equality we used (\ref{kx}) (i.e. $\pounds_k' x=0$)
and the fact that the Lie derivative commutes with the exterior differential.
Since this holds for any $W_1$, $W_2$ tangent to $S_{u_{0},v}$,
the only possibility is $d (k^{\prime}(y)) \propto dv$, i.e. that there exists
a function $G(v)$ on $\H_{u_0}$  such that $k^{\prime}(y) = G(v)$. Since
$k^{\prime}(y)$ vanishes on  a null generator $\gamma(v)$, it must be $G(v)=0$,
and hence $k^{\prime}(y)=0$. As the argument applies to each
$\H_{u}$, we conclude $k^{\prime \alpha} \nabla_{\alpha} y =0$, as claimed.

In summary, the vector field $k^{\prime}$ takes the form 
$$k^{\prime} = \frac{1}{L} \partial_v
$$ 
in the local coordinates $\{v,u,x,y\}$. Taking (\ref{kk}) into account and the definition of $x$
we derive the metric coefficients 
$g_{v\alpha} = - \frac{2L^2}{x^2} \delta_{\alpha}^{u} $ and the spacetime
metric must take the local form
\begin{align}
g = \frac{2L^2}{x^2} \left ( -2 du dv + dx^2 + dy^2 + 2 C_1 du dx
+ 2 C_2 du dy + C_{3} du^2 \right )
\label{metric}
\end{align}
for some functions $C_1,C_2,C_{3}$ depending on $u,x,y$ (because
$\xi = \partial_v$ is a Killing vector). The construction of the coordinates
described above was based on the choice of a transversal hypersurface $\Sigma$,
which in this coordinates becomes $\Sigma = \{ v=0 \}$. The freedom in 
choosing $\Sigma$ is reflected in the coordinate freedom $v = v' + s(u,x,y)$
which leaves the form of the metric invariant  and transforms the functions
$C_1$, $C_2$ and $C_{3}$ to
\begin{align*}
C'_{1} = C_1 - s_{,x} , \quad \quad
C'_{2} = C_2 - s_{,y} , \quad \quad
C'_{3} = C_{3} - 2 s_{,u}.
\end{align*}
Given the simple form of the metric we can now impose directly
the conditions of
$(\M,g)$ being $\Lambda$-vacuum and satisfying the alignment condition
(\ref{C=FF}). Computing the Ricci tensor of (\ref{metric}) one finds
\begin{align*}
R_{ux} - \Lambda g_{ux} & = \frac{1}{2} \partial_{y} \left ( \partial_x C_{2} - 
\partial_{y} C_1 \right ) = 0, \\
R_{uy} - \Lambda g_{uy} & = -\frac{x^2}{2} \partial_{x} \left [ x^{-2} \left (
\partial_x C_2 - \partial_{y} C_1  \right ) \right ] = 0. 
\end{align*}
The first equation implies that $\partial_x C_2 - \partial_y C_1 = q (x,u)$,
which inserted in the second yields $\partial_x ( x^{-2} q ) =0$, i.e.
$q = - x^2 r(u)$, for some smooth function $r(u)$. The functions $\tilde{C}_1$
and $\tilde{C}_2$ defined by $\tilde{C}_1 := C_1 - r(u) x^2 y$,
$\tilde{C}_2 = C_2$ satisfy then
\begin{align*}
\partial_x \tilde{C}_2 - \partial_y \tilde{C}_1 =
\partial_x C_2 - \partial_y C_1 + r(u) x^2 = 0
\end{align*}
so that there exists a function $s(u,x,y)$ such that $s_{,x}  =
\tilde{C}_1$ and $s_{,y}  = \tilde{C}_2$. Applying the coordinate
change $v = v' + s$ the form of the metric simplifies to (we drop the primes
in $v$ and set $M:= C'_{3}$)
\begin{align}
g = \frac{2L^2}{x^2} \left ( -2 du dv + dx^2 + dy^2 + 2 r(u) x^2 y du dx+
M du^2 \right ).
\label{metric2}
\end{align}
The $\Lambda$-vacuum field equations are satisfied if and only if $M$ solves
the linear inhomogeneous PDE
\begin{align*}
M_{,xx}  + M_{,yy} - \frac{2}{x} M_{,x} - r^2 x^4 =0.
\end{align*}
A particular solution of this equation is $\frac{1}{18} r^{2}(u) x^6$, so
that the general solution is
\be
M = H + \frac{1}{18} r^{2}(u) x^6 \label{MH}
\ee
with $H$ any solution of the homogeneous equation
\begin{align}
H_{,xx} + H_{,yy} - \frac{2}{x} H_{,x} =0. \label{eqH}
\end{align}
To solve this equation (cf. \cite{Si,P}),
perform the following change of dependent variable
\be
H=x^2\partial_x\left[\tilde h(u,x,y)/x \right] \label{Hh}
\ee
which transforms (\ref{eqH}) into 
$$
x\left(\tilde h_{,xxx} +\tilde h_{,yyx}\right)- \tilde h_{,xx}- \tilde h_{,yy}=
x\left( \tilde h_{,xx}+ \tilde h_{,yy}\right)_{,x}-\left(\tilde h_{,xx}+ \tilde h_{,yy}\right)=0
$$
whose general solution reads $\tilde h_{,xx}+ \tilde h_{,yy}=x f(u,y)$ for some smooth arbitrary function $f$ independent of $x$. This provides the general solution for $\tilde h$ as follows
$$
\tilde h = x g(u,y) + h(u,x,y)
$$
where $g(u,y)$ is arbitrary and 
\be
h_{,xx}+h_{,yy} =0. \label{Lap}
\ee
Introducing this into (\ref{Hh}) one checks that 
$$
H=x^{2} \left(g(u,y) + h(u,x,y)/x \right)_{,x}=x^{2} \left(h(u,x,y)/x \right)_{,x}
$$
so that $g(u,y)$ does not contribute to the function $H$ and the general solution of (\ref{eqH}) is given by 
\be
H=x^{2} \left(h/x \right)_{,x} \label{Hh2}
\ee
with $h(u,x,y)$ any solution of (\ref{Lap}).

The self-dual Killing form 
$\F_{\alpha\beta}$ associated to $\xi = \partial_v$ of the metric 
(\ref{metric2}) is
\begin{align*}
\F_{\alpha\beta} = k'_{\alpha} m_{\beta} - k'_{\beta} m_{\alpha}
\end{align*}
with $\bm{m} = \frac{L}{x} (dx + i dy)$ in agreement with 
(\ref{ss}). In particular $\F_{uv}=0$. Computing the self-dual Weyl
tensor of (\ref{metric2}) one finds 
$$
\C_{v \, u \, u \, y} = \frac{L^2 r(u)}{x}
$$
hence the alignment condition (\ref{C=FF}) forces $r(u)=0$. This
restriction turns out to be sufficient for the validity of 
(\ref{C=FF}), with  the
function $Q$ taking the value
\begin{align*}
Q = \frac{x^4}{8 L^2} \left ( - H_{,xx} + H_{,yy}  +2 i H_{,xy} \right )
\end{align*}
where $H$ is given by (\ref{Hh2}). 
In conclusion, the most general solution of $\Lambda$-vacuum spacetimes
with a Killing vector $\xi$ with null-self-dual Killing form 
and satisfying the alignment condition (\ref{C=FF}) can be written
in local form as
\begin{align}
g = \frac{2L^2}{x^2} \left ( -2 du dv + dx^2 + dy^2 + x^{2} \left(h/x \right)_{,x} du^2 \right ), \quad \quad h_{,xx} + h_{,yy}=0, 
\quad \quad \xi = \partial_v,
\label{metric3}
\end{align}
with $h$ any $u$-dependent solution of the Laplace equation (\ref{Lap}). Therefore, $h$ is the real part of any arbitrary $u$-dependent holomorphic function $\sigma(u,x+iy)$. 
This spacetime is known as the 
Siklos wave solutions \cite{Si} and corresponds also to the class $(IV)_0$ 
in \cite{ORR}, and they happen to be the only non-trivial Einstein spaces conformal to non-flat pp-waves \cite{Si}, see also \cite{GP,P}. 

It is noteworthy that anti-de Sitter (AdS) spacetime is included in the metric (\ref{metric3}) for the case with $h$ the real part of the function $\sigma =(A(u)-iB(u))\zeta^2$, giving 
$$
h=A(u)(x^2-y^2) +2B(u) x y \hspace{4mm} \Longrightarrow \hspace{4mm} H=A(u) (x^2+y^2).
$$
This case has $Q=0$, and the metric being conformally flat, it is a portion of AdS. Observe that the metric without the conformal factor $2L^2/x^2$ describes the electromagnetic plane waves \cite{Exact}, which are known to be conformally flat.

{\bf Remark:} Along the way, we have found another class of $\Lambda$-vacuum spacetimes given by (\ref{metric2}) with (\ref{MH}) together with (\ref{Lap}-\ref{Hh2}) and a {\em non-vanishing} function $r(u)$. This solution does not satisfy the special alignment condition (\ref{C=FF}) but, as one can easily check, it does satisfy the more general alignment condition (\ref{CF=F}). The Petrov type of this metric is \rm{III} and was identified previously in \cite{GDP} (case 3, formula (20) there).

\section{Main theorems and concluding remarks.}
\label{mainSect}
We finish the paper with a theorem that collects our main results
herein and with a discussion that summarizes all the results concerning characterizations of spacetimes subject to our main assumption (\ref{C=FF1}).

\begin{theorem}
\label{main}
Let $(\M,g)$  be a $\Lambda$-vacuum $C^3$ spacetime  admitting
a Killing vector $\xi$ with null self-dual
Killing form $\F_{\alpha\beta}$. Assume that the Weyl tensor
satisfies the alignment condition (\ref{C=FF}) where 
$Q$ is a $C^1$ function except
possibly at the boundary of the set
$\M^{\F} :=\{ \F_{\alpha\beta} = 0 \}$. 

If $\M^{\F} = \M$, then the spacetime is locally isometric to
the Minkowski flat spacetime and $\Lambda=0$.

If $\M^{\F} \neq \M$ then $\Lambda \leq 0$ and furthermore
\begin{itemize}
\item if $\Lambda <0$, then $\M^{\F} = \emptyset$ and
the spacetime is locally isometric to the 
{\bf Siklos wave spacetimes} (\ref{metric3}) ---including AdS as a particular case.
\item if $\Lambda=0$ the spacetime is either 
locally isometric to the {\bf vacuum plane waves}
(\ref{firstCase}) (when $\xi$ is
orthogonal to the wave vector of $\F_{\alpha\beta}$), or to the
{\bf vacuum and stationary Brinkmann spacetimes} (\ref{Stationary}) (when $\xi$ is not
orthogonal to the wave vector of $\F_{\alpha\beta}$). The intersection of both
cases is non-empty, corresponds to the case with $A$ and $B$ constant in 
the vacuum plane waves and leads to the {\bf irreducible Lorentzian locally
symmetric vacuum spacetimes} (\ref{LocSym1}).
\end{itemize}
\end{theorem}

As discussed in the introduction, one of the motivations of this work
was to complete the classification of $\Lambda$-vacuum spacetimes
admitting a Killing vector  and
satisfying the alignment condition (\ref{C=FF1}). When
$\F^2 \neq 0$ this was developed
in \cite{Mars1,Mars3} in the case of $\Lambda=0$ and in \cite{MS} 
when $\Lambda \neq 0$. In this paper we have 
considered the remaining case $\F^2=0$, so it makes sense to summarize
all these results here. 
Given that the entire class satisfying (\ref{C=FF1}) has now been fully identified we have now a clearer perspective on the results, and thus it is worth to re-consider and re-organize some of them. This is what we do next.

 The classification
in the case $\Lambda \neq 0$ and $\F^2 \neq 0$ discussed in \cite{MS}
leads to three
disjoint classes labeled (A), (B.i) and (B.ii) in Theorem 4 of \cite{MS},
and the Kerr-NUT-(A) de Sitter spacetime (which includes
the non-rotating Schwarzschild-NUT-(A) de Sitter case) is located partly
in the class (B.i) and partly in the class (B.ii). Something similar happens
in the vacuum case, but since the situation is simpler these two classes can
be reorganized so that they remain disjoint and one of them
contains the whole class of Kerr-NUT metrics (and their
plane and hyperbolic generalizations).
The following theorem provides this 
splitting in the vacuum $\F^2\neq 0$ case.
\begin{theorem}
\label{vacuumF2}
Let $(\M,g)$ be a smooth vacuum spacetime admitting a Killing vector
$\xi$ and satisfying the alignment condition (\ref{C=FF1}) where
$\F_{\alpha\beta}$ is the self-dual Killing form of $\xi$. Assume
that $\F^2 := \F_{\alpha\beta} \F^{\alpha\beta}$ is not identically zero.
Then the spacetime is locally isometric to 
an element of the following two mutually exclusive classes:
\begin{itemize}
\item {\bf gen-Kerr-NUT spacetime}. This class depends on 
two discrete constants $\sigma = \{ -1, 0 ,1 \}$ and $\delta = \{-1, 0, 1 \}$
and,
away from fixed points of Killing vectors of $(\M,g)$, there exist coordinates
$\{u,r,\theta,\varphi\}$ where $\xi = \partial_u$ and the metric is
\begin{align}
ds^2 = & - \frac{\Delta(r) + \Theta(\theta)}{\rho^2}
\left [ du + G(\theta) ( a G(\theta) + 2 \ell ) d \varphi 
\right ]^2 + \rho^2 \left ( d \theta^2 + G_{,\theta}{}^2 d \varphi^{2} \right ) \nonumber \\
& + 2 ( dr - a G_{,\theta}{}^2 d \varphi)
 \left [ du + G(\theta) ( a G(\theta) + 2 \ell ) d \varphi\right ] , 
\label{sigmaKerrNUT}
\\
\Delta(r) := & \sigma \left ( r^2 - \ell^2 \right ) - 2mr
-2 a \ell \alpha
, \quad \quad
\Theta(\theta) := a^2 G(\theta) \left ( \sigma G(\theta)- 2 \alpha 
\right ), \nonumber \\
\rho^2 := & r^2 + ( a G(\theta) + \ell)^2, \nonumber \\
G_{,\theta}{}^2 = & \delta +2 \alpha G(\theta)
- \sigma G^2(\theta), \quad \quad G_{,\theta} \mbox{ not identically zero}
\nonumber 
\end{align}
where $m,a,\ell,\alpha$ are arbitrary constants which, without loss
of generality, can be chosen to satisfy
$\alpha=0$ if $\sigma \neq 0$ and $\ell=0$ if $\{ \sigma=0,a \neq 0\}$. 
Moreover,
$\delta =1$ whenever $\{\sigma=1, \alpha=0\}$ or  $\{ \sigma =0, \alpha=0\}$.
\item {\bf Type D vacuum Kundt.} Away from fixed points of $\xi$, there
exists coordinates $\{t,u,r,x \}$, where $\xi = \partial_u$
and the metric is
\begin{align}
ds^2 & = (r^2 + \ell^2) \left ( - dt^2 +\Sigma_{,t}{}^2 dx^2 \right )
+ \frac{1}{W} dr^2 + W (du + 2 \ell \Sigma(t) dx)^2, \label{TypeD}  \\
W & = \frac{\sigma (r^2 - \ell^2) - 2 m r }{r^2 + \ell^2}, 
\quad \ell,m \in \mathbb{R}, \quad \quad
\Sigma_{,ttt} = \sigma \Sigma_{,t} \nonumber
\end{align}
where $\sigma = \{ +1,0,-1\}$.
\end{itemize}
\end{theorem}

\vspace{2mm}
{\bf Remark:} 
The name ``gen-Kerr-NUT spacetime'' is motivated by the fact that
the two-dimensional space at constant $r$ and $u$ is conformal
to a Riemannian surface of constant
curvature $\sigma$. When $\sigma=1$, this
class is the standard Kerr-NUT spacetime with mass $m$, specific
angular momentum $a$ and NUT charge $\ell$. The cases with $\sigma=0$ (resp.
$\sigma=-1$)
are analogous in the plane (resp. hyperbolic) cases. This notation follows the one used in 
Figure 16.2 in \cite{GP}.

\proof Setting $\Lambda=0$ in Theorem 4 in \cite{MS} we find that case $(A)$
is impossible. The local metric 
in case (B.i) is (away from fixed points of Killing vectors of $(\M,g)$)
\begin{align}
ds^2 & = - N \left ( dv - Z^2 dx \right )^2
+ 2 \left ( dy + V dx \right ) \left ( dv - Z^2 dx \right ) 
+ (y^2 + Z^2) \left ( \frac{dZ^2}{V} + V dx^2 \right ) \label{Pleban} \\
  \xi & = \partial_v, \quad N = c 
- \frac{b_1 y + b_2 Z}{y^2 + Z^2}, 
\quad V =k +b_2 Z - c Z^2.
\end{align}
where $k,b_1,b_2,c$ are constants with the property that $V(Z)>0$
in some non-empty interval where $Z$ takes values.
Define $\s>0$ and $m$ by
$$c = \sigma \s^2 , \hspace{3mm} b_1= 2 m\s^3
$$
(when $c=0$, $\s$ can be fixed to any non-zero value) and 
introduce three constants $\ell,\alpha,a$, with $a>0$
as a solution of the under-determined system
\begin{equation}
\left . \begin{array}{ll}
        b_2 = 2 \s^3 ( \sigma \ell + a \alpha ) \\
        k = \s^4 ( -\sigma \ell^2 - 2 a \alpha \ell + \delta a^2  )
        \end{array}
\right \}. \label{sys}
\end{equation}
To show that this system is compatible
note that 
\begin{itemize}
\item when $\sigma=\pm 1$
we can choose $\{2 \ell = \sigma b_2 \s^{-3}, \alpha=0 \}$ providing $4\delta a^{2}= 4ks^{-4}+\sigma b_{2}^{2}s^{-2}$ (and $a > 0$
arbitrary if, in addition, $4ks^{-4}+\sigma b_{2}^{2}s^{-2}=0$ requiring $\delta =0$),
\item and when
$\sigma=0$ we can choose $\{ \ell=0, 2\alpha = b_2 a^{-1} \s^{-3}\}$ 
where $a$ is given by $\delta a^{2}=ks^{-4}$ (or 
$a > 0$ arbitrary if $k=0$ which requires $\delta =0$). 
\end{itemize}
Nevertheless all the expressions below
hold for any solution of (\ref{sys}).
Consider the coordinate change $Z = \s ( a G(\theta) + \ell)$
in terms of which the polynomial $V(Z)$ becomes
\begin{align*}
V(Z(\theta)) = \s^4 a^2 \left ( \delta + 2 \alpha G(\theta)
- \sigma G^2(\theta) \right ) > 0
\end{align*}
(this shows in particular that $\delta=1$ if 
$\{\sigma=1, \alpha=0\}$ or if $\{ \sigma=0,\alpha=0\}$, as claimed)
and fix $G(\theta)$ 
 by requiring that $dZ^2/V =s^{-2}d\theta^2$, that is to say, as any solution of the ODE
\begin{align}
G_{,\theta}{}^2 &= \delta + 2 \alpha G(\theta) - \sigma G^2(\theta) , 
\quad \quad G_{,\theta} \mbox{ not identically zero} \label{eq1}.
\end{align}
The condition $G(\theta)$ not constant is necessary for $Z(\theta)$
to define a coordinate change. Note that, in these circumstances 
$G(\theta)$ also solves
\begin{align}
G_{,\theta\theta} & = \alpha - \sigma G(\theta). \label{eq2}
\end{align}
With the additional coordinate changes 
\begin{align*}
y = \s r, \quad  v = \frac{1}{\s}  (u - \frac{\ell^2}{ a} \varphi), \quad 
x= - \frac{1}{a \s^3} \varphi
\end{align*}
a straightforward calculation brings the metric (\ref{Pleban}) into the
form (\ref{sigmaKerrNUT}) (note that in this process the Killing vector
field $\xi$ needs to be rescaled appropriately).

Concerning Case (B.ii) with $\epsilon = 1$ in Theorem 4 of \cite{MS},
the spacetime metric (away from fixed points of $\xi$) is
\begin{align}
ds^2 & = - W \left ( dv - \hat{\bm{w}} \right )^2 + 2 dy \left ( dv - 
\bm{\hat{w}} \right ) + (\beta^2+ y^2) h_{+}  \label{metricBii1} \\
W & = (\beta^2 + y^2)^{-1}   \left ( 
 - \kappa
\left ( \beta^2 - y^2 \right ) + n y  \right ), \quad \quad
\hat{d} \hat{\bm{w}} 
= 2 \beta \bm{\eta_{+}}  
\quad \quad 
\xi  = \partial_v,  \label{Eqw}
\end{align}
where $h_{+}$ is a positive definite metric of constant curvature
$\kappa$ and volume form $\bm{\eta_{+}}$
and $\kappa, \beta,n$ are real constants such that $W(y)$
is positive in some interval where $y$ takes values.
Define $\s>0$, $\ell$ and $m$ by 
$$
\kappa = \sigma \s^2 ,Ê\hspace{5mm} \beta = -\ell \s , \hspace{5mm}  n = - 2 m \s^3 .
$$
Being $h_+$ of constant
curvature $\sigma \s^2$ it can be written in local form as
$$
h_{+} = \frac{1}{\s^2} \left ( d \theta^2 + G_{,\theta}{}^2 d \varphi^2
\right )
$$
where $G(\theta)$ is a non-constant function
satisfying the equation 
$G_{,\theta\theta\theta} = - \sigma G_{,\theta}$ . In fact, we can assume
without loss of generality that $G(\theta)$ satisfies
(\ref{eq1}) (and hence also (\ref{eq2})). Equation (\ref{Eqw}) for 
$\hat{\bm{w}}$ can be integrated explicitly as
$\hat{\bm{w}} = - 2 \ell \s^{-1} G(\theta) d \varphi + df$, where $f$
is any function of $\{ \theta,\varphi \}$. Performing the coordinate
change
\begin{align*}
y = \s r, \quad \quad v = \frac{1}{\s} u + f,
\end{align*}
the metric  (\ref{metricBii1}) becomes
\begin{align*}
ds^2 = & \frac{\sigma (r^2 - \ell^2) - 2 m r}{r^2 + \ell^2} 
( du + 2 \ell G(\theta) d \varphi)^2
+ 2 dr ( du + 2 \ell G(\theta) d \varphi) \\
& + (r^2 + \ell^2 )( d \theta^2 + G_{,\theta}{}^2 d \varphi^2 )
\end{align*}
which corresponds to the gen-Kerr-NUT class with
vanishing $a$.

Finally, case (B.ii) with $\epsilon = -1$ in Theorem
4 of \cite{MS} corresponds to the metric
\begin{align*}
ds^2 & = W^{-1} dy^2 + W \left ( dv - \hat{\bm{w}} \right )^2 +
(\beta^2 + y^2 ) h_{-}   \\
W & = (\beta^2 + y^2)^{-1}   \left ( 
 - \kappa \left ( \beta^2 - y^2 \right ) + n y  \right ), \quad \quad
\hat{d} \hat{\bm{w}} 
= 2 \beta \bm{\eta_{-}} 
\quad \quad 
\xi  = \partial_v,  
\end{align*}
where $h_{-}$ is a Lorentzian metric of constant curvature
$\kappa$ and volume form $\bm{\eta_{-}}$
and, as before, $\kappa, \beta,n$ are real constants such that $W(y)$
is positive in some interval where $y$ takes values. We proceed
similarly. Define $\s$ by $\kappa = \sigma \s^2$ and
choose local coordinates $\{ t, x\}$
so that 
$$
h_{-} = \s^{-2} (-dt^2 + \Sigma^2_{,t}(t) dx^2 ),
$$
The condition that 
this metric has constant curvature $\sigma \s^2$ is $\Sigma_{,ttt}= \sigma \Sigma_{,t}$.
Define $\ell, m$ by $\beta = - \s \ell$ and $n= -2 m \s^3$.
The equation for  $\bm{\hat{\omega}}$ can be integrated as
$\bm{\hat{\omega}} = -2 \s^{-1} \ell H dx + f(t,x)$.
The coordinate change $\{ v = \s^{-1} u + df, y=\s r\}$
brings the metric into the 
from (\ref{TypeD}). \hfill $\Box$

\vspace{2mm}
{\bf Remark.} 
As shown in Lemma 4 in \cite{Mars1}, spacetimes satisfying the hypotheses of Theorem  \ref{vacuumF2}
admit an exact
 Ernst one-form $\bm{\chi}$ associated to $\xi$, which defines the Ernst potential $\chi$ by $\bm{\chi}=d\chi$. Moreover, selecting
(partially) the free additive constant in $\chi$ so that
$\mbox{Re} (\chi) = - g(\xi,\xi)$, there exist complex constants
$c$ and $A$ such that 
%
%
\begin{align*}
Q = \frac{-6}{c-\chi}, \quad \quad \F^2 = A (c - \chi)^4.
\end{align*}
The constants $c$ and $A$ play an important role in characterizing
locally the Kerr metric 
among vacuum spacetimes admitting a Killing vector $\xi$
such that the alignment condition (\ref{C=FF}) holds
and $\F^2 \neq 0$ somewhere. Such a purely local characterization was first
given in Theorem 1 in  \cite{Mars2}, where it was claimed that
the Kerr spacetime can be characterized by the condition
$ \mbox{Re}(c) > 0$
and $A$ real and negative.
Unfortunately, Theorem 1 in \cite{Mars2} is incorrect
as stated because, as we show below, the type D vacuum
Kundt spacetime also admits  a particular case with $\mbox{Re} (c) > 0$ and
$A<0$. The reason why this possibility was missed in Theorem 1 in \cite{Mars2}
was that the arguments in \cite{Mars1} (on which the validity 
of Theorem 1 was based)  made the implicit assumption
that the Killing vector field $\xi$ is
not everywhere orthogonal to the
plane generated by the two real eigenvectors $\{ l, k \}$ of the self-dual
Killing form
$\F_{\alpha\beta}$ (this condition is automatically satisfied if
$\xi$ is timelike somewhere, which was the case of interest in
\cite{Mars1}).
This assumption  needs to be added to the conditions on $c$ and
$A$ in the local characterization of Kerr in order to make the result
correct. We 
provide here a corrected version of the theorem and give its proof.
\begin{theorem}[Corrected local characterization of the
Kerr metric]
\label{corrected}
Let $(\M,g)$ be a vacuum spacetime admitting a Killing vector
$\xi$ such that the alignment condition (\ref{C=FF}) holds
with $\F^2 \neq 0$ on at least one point. Then there exist complex
constants $c,A$ such that 
\begin{align*}
Q = \frac{-6}{c-\chi}, \quad \quad \F^2 = A (c - \chi)^4,
\end{align*}
where $\chi$ is the Ernst potential (i.e. a function satisfying 
$\nabla_{\beta} \chi = 2 \xi^{\alpha} \F_{\alpha\beta}$ and which 
necessarily exists globally on $(\M,g)$).  Fix partially the 
free additive constant in $\chi$ so that $\mbox{Re} (\chi) = -g (\xi,\xi)$.
Then $(\M,g)$ is locally isometric to a Kerr spacetime if and only if
the following two conditions are satisfied
\begin{itemize}
\item[(i)] $\mbox{Re} (c) > 0$ and $A$ is real and negative.
\item[(ii)] There is at least one point $q \in \M$ where
$\F^2 |_q \neq 0$ such that $\xi|_q$ is not orthogonal to the $2$-plane 
generated by the two
real null eigenvectors  $\{ l|_q, k|_q \}$ of $\F_{\alpha\beta}|_q$.
\end{itemize}
\end{theorem}
\proof As discussed above, the ``if'' part follows from the arguments
in \cite{Mars1}. For the ``only if'' part we could simply compute the vectors
$\{ k, l \}$ and the constants $c$ and $A$
in the Kerr spacetime and show that both (i) and (ii) hold.
However, for completeness we compute these objects for
the full class of spacetimes contained in Theorem \ref{vacuumF2}
and give a direct and independent proof of the theorem.

For the gen-Kerr-NUT spacetime (\ref{sigmaKerrNUT}) with the given choice of $\xi =\partial_u$, a direct calculation shows that
the  Ernst potential satisfying $\mbox{Re} (\chi) = - g(\xi,\xi)$
is 
\begin{align*}
\chi & = \frac{\Delta (r)+ \Theta (\theta)+ 2 i  \left [ 
(a \alpha + \sigma \ell ) r - m ( a G(\theta) + \ell) \right ] }{\rho^2} 
+ i \omega_0
\end{align*}
where $\omega_0 \in \mathbb{R}$ is any constant. Moreover, $c$
and $A$ are given by
\begin{align*}
c & = \sigma + i \omega_0, \quad \quad 
A = \frac{-1}{4 \left[m - i ( a \alpha + \sigma \ell )\right]^2}.
\end{align*}
The null eigendirection of $\F_{\alpha\beta}$ with eigenvalue
$(c- \chi)^2/[4 ( m - i (a \alpha + \sigma \ell)]$
is generated by $k = \partial_r$.
Hence $g(\xi,k)  = g (\partial_u,\partial_r) =1$
and $\xi$ is not orthogonal to the $2$-plane generated by $\{ l,k\}$.

For the Type D Kundt metric, the Ernst potential, the complex constants
$c$ and $A$, and the real null eigenvectors $\{ l, k \}$ of $\F_{\alpha\beta}$ are 
\begin{align*}
\chi & = \frac{- \sigma (r^2 - \ell^2 ) + 2 m r + 2 i \ell (r \sigma - m)
 }{r^2 + \ell^2} + i \omega_0 \quad \quad
c  = - \sigma + i \omega_0, \quad \quad A = \frac{1}{4 (m + i \sigma \ell )^2}, \\
\bm{l} & = dt + \Sigma_{,t} dx \quad 
\mbox{ with eigenvalue } \frac{-i (\sigma + \chi)^2}{4
(m+ i \sigma l)}, \\
\bm{k} & = dt - \Sigma_{,t} dx \quad 
\mbox{ with eigenvalue } \frac{i (\sigma + \chi)^2}{4
(m+ i \sigma l)}.
\end{align*}
The eigenvectors are given in terms of their metrically associated one-forms
$\{ \bm{l}$, $\bm{k}\}$
so that it is obvious that $\xi = \partial_u$ is everywhere orthogonal
to the two-plane generated by $\{ l,k \}$.

The Kerr spacetime corresponds precisely to the subclass with $\sigma = +1$
and NUT parameter $\ell=0$ of the gen-Kerr-NUT class. 
As $\sigma =1$ from Theorem \ref{vacuumF2} we know that we can set $\alpha =0$ without loss of generality and then $A$ is real and negative.
However, the subclass with $\{ m=0, \ell \neq 0, \sigma =-1\}$
in the type D vacuum Kundt class also has the properties
that $\mbox{Re}(c) >0$
and $A<0$  and hence satisfies hypothesis (i) in  the theorem.
From the properties of the null eigenvectors $\{ l,k\}$ 
of these spacetimes it is clear that
item (ii) selects uniquely the Kerr class, as claimed.
\hfill $\Box$

\vspace{3mm}

For completeness, we write down
the second null eigendirection of $\F_{\alpha\beta}$. It has eigenvalue
$-(c- \chi)^2/[4 ( m - i (a \alpha + \sigma \ell)]$
and it is generated by
\begin{align*}
l = \partial_{u} + \frac{a^2 \delta + \Delta(r)}{2(r^2 + \ell^2)}  
\partial_r 
+ \frac{a}{r^2 + \ell^2}   \partial_{\varphi}.
\end{align*}

\vspace{5mm}

We conclude the paper
with Table \ref{Summary} which summarizes  
the results in this paper and those in \cite{Mars1,Mars2,MS}.

\begin{table}[!ht]
\begin{center}
\begin{tabular}{M{12mm}|M{26mm}M{22mm}|P{\cellwidth}|}
& \multicolumn{2}{|c|}{$\F^2=0$} & $\F^2 \neq 0$ \\
\hline
\multirow{6}{*}{$\Lambda \neq 0$}  & & &
\multirow{6}{*}{
\specialcell[t]{
- Pleba\'nski (Kerr-(A) de Sitter $a \neq 0$...). \\
- Pure LRS (Schwarzschild-(A) de Sitter, Taub-NUT-(A) de Sitter, ...). \\
- Type D Kundt. \\
- Product metrics (uncharged Bertotti-Robinson, Nariai). \\
- dS and AdS.
}}
\\
& $\Lambda > 0$ & $\nexists$ & \\
& & & \\
\cline{2-3}
& & & \\
& $\Lambda<0$ & Siklos (AdS). & \\
&  &  & \\
\hline
& & & \\
$\Lambda =0$
& 
\specialcell{
- Vacuum plane waves. \\
- Vac. stationary Brinkmann. \\
(vac. irreducible loc. sym., \\
Minkowski)} &  & gen-Kerr-NUT, \hspace{5mm} Vacuum Type D Kundt (Minkowski) \\
& & & \\
\hline
\end{tabular}
\caption{The column $\F^2 =0$ is the content of Theorem \ref{main}
and we have emphasized that (a) AdS is included in the Siklos class and (b) that the vacuum plane waves and the {\em vacuum} stationary Brinkmann
classes are not disjoint by including (in parenthesis) their
intersection, namely the irreducible locally symmetric vacuum spacetimes.
The cell $\F^2 \neq 0, \Lambda=0$ is the content of Theorem \ref{vacuumF2}.
The cell $\F^2 \neq 0, \Lambda \neq 0$ was proven in Theorem 4
in \cite{MS}. The Pleba\'nski case corresponds to case (B.i) in that Theorem, while its
case (B.ii) with $\epsilon = 1$ corresponds to pure LRS metrics:
these are metrics admitting a four-dimensional local
isometry group acting on three-dimensional hypersurfaces. These
spacetimes were classified by Cahen and Defrise \cite{CD} and contain many
special subcases, some of which are explicitly listed in parenthesis. The
type D Kundt metrics correspond to case (B.ii) with $\epsilon = -1$ in Theorem 4
of \cite{MS}, and they contain both cases with $\Lambda$ vanishing or not.
The Kerr-NUT-(A) de Sitter spacetime belongs to the Pleba\'nski class
as long as the rotation parameter $a$ is non-zero. The case
of vanishing $a$ belongs to the pure LRS case. 
Finally the ``product metrics'' correspond to case
(A) in \cite{MS} for which the spacetime metric
is locally the product of a Lorentzian two-dimensional metric times a Riemannian two-dimensional
metric, both of them of constant curvature $\Lambda$. As a final  observation, it is noteworthy that de Sitter, Minkowski and anti-de Sitter spacetimes are included here, but while AdS and Minkowski have Killing vector fields with $\F^2$ both vanishing or not, dS does not admit any Killing vector with null Killing 2-form.} 
\label{Summary}
\end{center}
\end{table}

\vspace{1.2em}

\noindent {\textbf {Acknowledgements}}

\vspace{0.5em}

MM acknowledges financial support under projects  FIS2012-30926 and
FIS2015-65140-P (Spanish
MINECO). JMMS is supported by grants FIS2014-57956-P (Spanish MINECO-fondos FEDER),
and UFI 11/55 (UPV/EHU).

\end{document}